\documentclass[aps,prb,twocolumn,floats]{revtex4}
\usepackage{amssymb}
\usepackage{amsbsy}
\usepackage{amsmath}
\usepackage{graphicx}
\usepackage{amsmath}
\usepackage{times}
\usepackage{color}
\usepackage{subfigure}
\usepackage{setspace}
\usepackage{bm}

\newcommand\bea{\begin{eqnarray}}
\newcommand\eea{\end{eqnarray}}
\newcommand\beq{\begin{equation}}
\newcommand\eeq{\end{equation}}
\newcommand\bib{\bibitem}

\newcommand{\noi}{\noindent}
\newcommand{\non}{\nonumber}
\newcommand{\al}{\alpha}
\newcommand{\be}{\beta}
\newcommand{\ga}{\gamma}
\newcommand{\Ga}{\Gamma}
\newcommand{\de}{\delta}
\newcommand{\De}{\Delta}
\newcommand{\om}{\omega}
\newcommand{\si}{\sigma}

\newcommand{\da}{\dagger}

\newcommand{\vk}{{\vec k}}

\begin{document}

\title{Low-frequency phase diagram of irradiated graphene and periodically
driven spin-1/2 $XY$ chain}

\author{Bhaskar Mukherjee$^1$, Priyanka Mohan$^2$, Diptiman Sen$^3$, and
K. Sengupta$^1$}

\affiliation{$^1$Theoretical Physics Department, Indian Association for the
Cultivation of Science, Jadavpur, Kolkata 700032, India \\
$^2$Harish-Chandra Research Institute, HBNI, Chhatnag Road, Jhunsi,
Allahabad
211019, India \\
$^3$Center for High Energy Physics, Indian Institute of Science, Bengaluru,
560012, India}

\date{\today}

\begin{abstract}

We study the Floquet phase diagram of two-dimensional Dirac
materials such as graphene and the one-dimensional (1D) spin-1/2
$XY$ model in a transverse field in the presence of periodic
time-varying terms in their Hamiltonians in the low drive frequency
($\om$) regime where standard $1/\om$ perturbative expansions fail.
For graphene, such periodic time dependent terms are generated via
the application of external radiation of amplitude $A_0$ and time
period $T = 2\pi/\om$, while for the 1D $XY$ model, they result from
a two-rate drive protocol with time-dependent magnetic field and
nearest-neighbor couplings between the spins. Using the
adiabatic-impulse method, whose predictions agree almost exactly
with the corresponding numerical results in the low-frequency
regime, we provide several semi-analytic criteria for the occurrence
of changes in the topology of the phase bands (eigenstates of the
evolution operator $U$) of such systems. For irradiated graphene, we
point out the role of the symmetries of the instantaneous
Hamiltonian $H(t)$ and the evolution operator $U$ behind such
topology changes. Our analysis reveals that at low frequencies,
topology changes of irradiated graphene phase bands may also happen
at $t= T/3, 2T/3$ (apart from $t=T$) showing the necessity of
analyzing the phase bands of the system for obtaining its phase
diagrams. We chart out the phase diagrams at $t=T/3, 2T/3,\, {\rm
and }\, T$, where such topology changes occur, as a function of
$A_0$ and $T$ using exact numerics, and compare them with the
prediction of the adiabatic-impulse method. We show that several
characteristics of these phase diagrams can be analytically
understood from results obtained using the adiabatic-impulse method
and point out the crucial contribution of the high-symmetry points
in the graphene Brillouin zone to these diagrams. We study the modes
which can appear at the edges of a finite-width strip of graphene
and show that the change in the number of such modes agrees with the
change in the Chern number of bulk graphene as we go across a phase
band crossing. Finally we study the 1D $XY$ model with a two-rate
driving protocol. After studying the symmetries of the system, we
use the adiabatic-impulse method and exact numerics to study its
phase band crossing which occurs at $t=T/2$ and $k=\pi/2$. We also
study the end modes generated by such a drive and show that there
can be anomalous modes whose Floquet eigenvalues are not equal to
$\pm 1$. We suggest experiments to test our theory.

\end{abstract}


\maketitle

\section{Introduction}
\label{intro}

The physics of closed quantum systems driven out of equilibrium has
attracted a lot of theoretical interest in recent times
\cite{rev1,rev2,rev3,rev4}. Such dynamics becomes particularly
interesting during the passage of the system through a quantum
critical point where it becomes non-adiabatic. The excess energy
$\de E$ and density of excitations $n$ through such a critical point
obey, for slow, linear or non-linear, power-law quenches, universal
scaling laws \cite{kz1,pol2,diva1,ks1,ks2,pol3,sondhi1}. More
recently, analogous scaling laws have also been derived for fast
quenches \cite{sdas1,sdas2}. The properties of such driven systems
have also been studied following a sudden quench where a parameter
of the system is changed instantly. Such sudden quenches leads to
transient oscillations whose amplitude peaks when the final
Hamiltonian is near a critical point \cite{subir1,das1}; in addition
they lead to interesting steady states \cite{cala1}. Moreover, the
work statistics of such systems is tied to the Locschmidt echo and
may display edge singularities \cite{silva1}. Such
out-of-equilibrium dynamics also leads to dynamical transitions
which have no counterparts in equilibrium systems
\cite{dytr0,dytr1,dytr2,amit1}. More recently, such studies has been
extended to cases where two parameters of the system Hamiltonian are
varied as functions of time with different rates; such dynamics
leads to a generalization of the well-known Kibble-Zurek scaling
laws \cite{sau1} and provide a route to realization of quantum
dynamics with controlled fidelity \cite{del1}. Furthermore, there
have been several studies on the applicability of renormalization
group methods for such out-of-equilibrium systems; such studies are
expected to shed light on the possibility extending the concept of
universality to such driven systems \cite{rg1,rg2,rg3}. The
motivation for such theoretical studies has received experimental
support from ultracold atom systems \cite{rev2}. The isolated nature
of such systems makes them perfect test beds for studying coherent
quantum dynamics of closed non-equilibrium systems
\cite{exp1a,exp1b}.

More recently, a significant emphasis in theoretical studies of
driven closed quantum systems has shifted to systems driven out of
equilibrium using a periodic protocol. The chief reason for this
stems from the recognition that such drives lead to a host of
interesting phenomena which have no counterparts in aperiodically
driven systems. For example, periodically driven systems exhibit
Stuckelberg interference phenomenon \cite{stu1}; the signature of
such interference phenomenon leads to experimentally discernible
features in their excitation densities and the statistics of work
distribution \cite{stu2,stu3}. Moreover, such driven integrable
systems exhibit a separate class of dynamical transition which
leaves its signature on local correlation functions \cite{asen1}. In
addition, they also exhibit dynamic freezing at specific
frequencies; at these frequencies the wave function of the system
after a single or multiple drive period(s) exhibits a near unity
overlap with the initial wave function \cite{adas1,sm1,sk1}.
Furthermore, such drives may also lead to novel steady states which
do not have any aperiodic counterparts \cite{adas2}.

An aspect of periodically driven closed quantum systems that has
gained recent attention involves a change in topology and the
concomitant generation of edge modes of these systems as a function
of the drive frequency \cite{top1,top2}. This phenomenon has been
mostly studied either in the context of graphene or topological
insulators (whose low-energy quasiparticles obey a Dirac-like
equation) in the presence of circularly polarized light
\cite{oka1,gil1} or for model Hamiltonians with engineered drive
protocols which allow for a simple analysis of the phenomenon
\cite{top1}. A central role in such studies is played by the
time-evolution operator of the driven system which can be expressed,
in terms of its Hamiltonian $H (t)$, as \bea U(t,0) &=& {\mathcal
T}_t e^{- i \int_0^t dt' H(t')/\hbar}\label{evol1}, \eea where
${\mathcal T}_t$ denotes time ordering. The Floquet Hamiltonian
$H_F$, which describes the properties of the driven system at the
end of an integer number of drive periods can be read off from $U$
using the definition \bea U(n T,0) &=& e^{-i n H_F T/\hbar}, \quad n
\in Z. \label{floq1} \eea Initial works on the subject analyzed such
changes of topology via a study of the properties of $H_F$
\cite{abhiskar1}. However, it was later found that the study of the
Floquet Hamiltonian, which amounts to a stroboscopic tracking of the
time evolution after an integer number of time periods, is not
always adequate for this purpose \cite{top1}. Instead, it is
sometimes necessary to track the time evolution of the phase bands,
{\it i.e.}, the eigenvalues $U(t,0)$, which control the dynamics of
the system within a single time period $T$. The crossing of such
phase bands has been shown to be intimately tied to the change in
the topology of the driven system \cite{top1}. The precise
conditions for the occurrence of such phase band crossings leading
to a change in the topology of the system and the generation of edge
modes has been charted out for cases where a single parameter of the
system Hamiltonian is driven periodically \cite{bm1}. However such
an analysis has not been extended to two-rate protocols, {\it i.e.},
to situations where two parameters of the system Hamiltonian are
driven periodically.

Examples of the latter class of driven systems include graphene in
the presence of circularly polarized external radiation as studied
in Refs.\ \onlinecite{oka1} and \onlinecite{gil1}. We note here that
the high-frequency phase diagram of irradiated graphene has been
studied analytically in terms of its Floquet Hamiltonian using
several expansion techniques all of which relies on the smallness of
$1/\om = T/(2\pi)$ \cite{oka1,gil1,kobo1,kundu1}. However, to the
best of our knowledge, there is no analytic method which allows a
systematic understanding of the low-frequency phase diagram of
irradiated graphene; in particular, the role of symmetries of the
instantaneous Hamiltonian and the unitary evolution operator behind
the change in the topology of these driven systems have not been
analyzed so far in this regime. Moreover, the phase bands of
graphene-like Hamiltonians hosting Dirac quasiparticles in the
presence of external radiation has also not been studied so far in
the low drive frequency regime.

In this work, we analyze the phase bands of two integrable models
subjected to periodic drives. The first of these is the
two-dimensional (2D) Dirac Hamiltonian of graphene in the presence
of circularly polarized external radiation of amplitude $A$ and
frequency $\om=2 \pi/T$, while the second is the 1D spin-1/2 $XY$
Hamiltonian with periodically varying nearest-neighbor interactions
and in the presence of a time-periodic magnetic field. The main
results that we obtain from our study are as follows. First, we
develop an adiabatic-impulse method which provide a near-exact match
with results obtained via exact numerics at low $\om$ and allows us
to obtain semi-analytic expressions for the phase bands of these
models in the low-frequency regime where standard perturbative
$1/\om$ expansions fail. Second, for irradiated graphene, using
general symmetry analysis of both the instantaneous Hamiltonian
$H(t)$ and the evolution operator $U$, we show that the change in
topology of the phase bands of irradiated graphene are most likely
to occur at special high symmetry points in its Brillouin zone;
moreover, such crossings can generically occur at $t=T/3, 2T/3, \,
{\rm and}\, T$ showing the inadequacy of Floquet Hamiltonian based
analysis and the necessity of the use of phase bands for analyzing
such systems at low radiation frequencies. Third, we provide a set
of semi-analytic conditions necessary for the phase bands of
irradiated graphene to cross. Since such crossings are responsible
for change of topology of the state of the system, our method
provides a semi-analytic understanding of the phase diagram
irradiated in the low-frequency regime where standard high-frequency
expansion techniques fail. Fourth, we chart out the phase diagram
for irradiated graphene at $t=T/3, 2T/3, \, {\rm and}\, T$, discuss
their features, and compare them with that obtained using
adiabatic-impulse method. We show that several characteristics of
these phase diagrams may be analytically understood using
predictions of the adiabatic-impulse method. Fifth, we obtain the
conditions for phase band crossings at $t=T/2$ in the 1D $XY$ model
with a two-rate protocol, {\it i.e.}, with periodically time-varying
couplings and magnetic field with two different drive frequencies.
We find that such crossings occur at $k=\pi/2$ and have no analog in
single rate drive protocols studied earlier \cite{bm1}. We also show
that such driven XY chains support end modes including anomalous
modes whose Floquet eigenvalues are not equal to $\pm 1$. Finally,
we discuss realistic experiments which may test our theory.

The plan of this paper is as follows. In Sec.\ \ref{secsymm}, we
chart out the symmetries of the instantaneous Hamiltonian and the
evolution operator describing graphene under external radiation.
This is followed by Sec.\ \ref{secadimp}, where we present details
of the adiabatic-impulse approximation and chart out the generic
conditions for topology change of driven systems based on this
approximation. Next, in Sec.\ \ref{secphd}, we compare the results
obtained from the adiabatic-impulse approximation with exact
numerics, chart out semi-analytic conditions for phase band
crossings, and present the phase diagram of irradiated graphene
obtained using exact numerics and adiabatic-impulse approximation
method. This is followed by Sec.\ \ref{sec1dxy} where we study the
driven 1D $XY$ model and discuss its phase band crossings and end
modes. Finally we summarize our results, discuss their experimental
implications, and conclude in Sec.\ \ref{secdiss}.

\begin{figure}[t!]
\begin{center}
\includegraphics[width=\columnwidth]{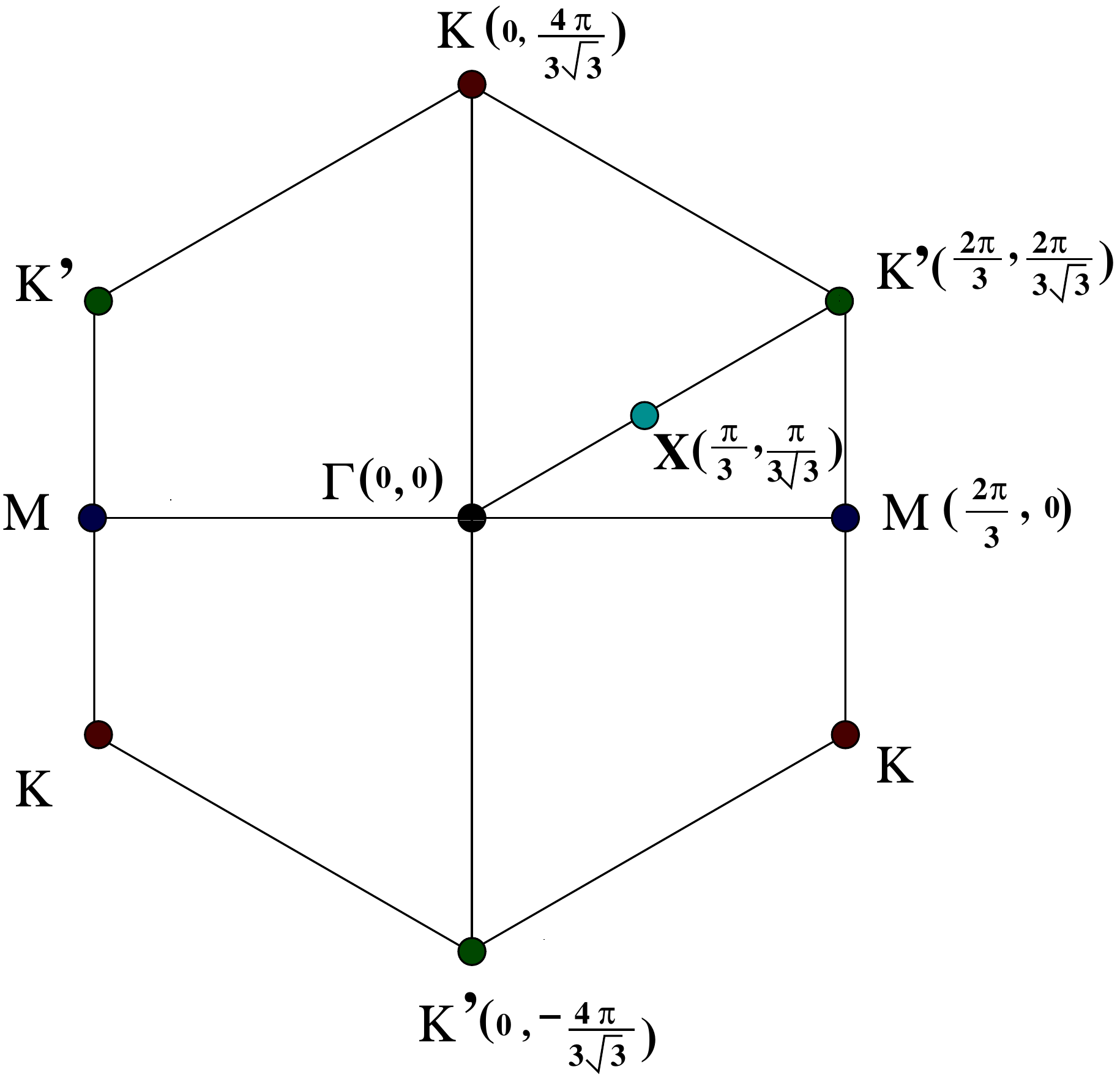}
\end{center}
\caption{Brillouin zone of graphene showing the $\Ga$ point, the
Dirac points $K$ and $K'$, the $M$ points and the $X$ point. Out of
these the first three ($\Gamma$, $K,\, K'$ and $M$) hosts maximal
number of phase band crossings.} \label{fig1}
\end{figure}

\section {Symmetries of the graphene Hamiltonian with external radiation}
\label{secsymm}

In this section, we analyze the symmetry properties of both the Hamiltonian
of graphene in the presence of external radiation and the corresponding
evolution operator defined in Eq.\ \eqref{evol1}. To this end, we take the
nearest-neighbor spacing in graphene to be $a_0$ and the hopping amplitude to
be $-\ga$. (We shall henceforth set $\ga$ and $a_0$ equal to 1 unless
mentioned otherwise). Given a site on the $A$ sublattice, we choose the
vectors to its three nearest-neighbor sites on the $B$ sublattice to be
\bea {\vec a}_{1[2]} &=& (1/2,+[-]\sqrt{3}/2),\quad {\vec a}_3 = (-1,0).
\label{vecsite} \eea
The reciprocal lattice vectors are
\bea {\vec G}_1 &=& (2 \pi/3, 2\pi /\sqrt{3}), \quad {\vec G}_2 = (2 \pi
/3, -2 \pi/\sqrt{3}). \label{recvec} \eea
We will choose the Brillouin zone to be a
hexagon as shown in Fig.\ \ref{fig1}.

We now apply circularly polarized electromagnetic radiation to
graphene. We take the vector potential ${\vec A}=(A_x,A_y)$ to be of
the form $A_x = A_0 \cos (\om t)$ and $A_y = A_0 \sin (\om t)$,
where $A_0$ and $\om = 2\pi/T$ are respectively the strength and
frequency of the radiation and $T$ is its time period. Following the
Peierls prescription, the vector potential is incorporated into the
hopping amplitude as phases given by $- (e /c) {\vec a}_j \cdot
{\vec A}$ on a bond labeled as $j$, where $-e$ is the charge of the
electron and $c$ is the speed of light. Let us define the parameter
$\al = e A_0/c$. In momentum space, the hopping amplitude on bond
$j$ takes the form $-\ga \exp [i {\vec a}_j \cdot ({\vec k} + \al
{\vec A})]$. The Hamiltonian of graphene then takes the form
\bea && H_G (t) ~=~ \sum_\vk ~\psi_\vk^\da H_\vk (t) \psi_\vk, \non \\
&& H_\vk (t) ~=~ Z_\vk (t) \tau_+ + Z^{\ast}_\vk (t) \tau_-, \non \\
&& Z_\vk (t) ~=~ - e^{ i (\al \cos(\om t-\pi/3) + (k_x +
\sqrt{3} k_y)/2) } \non \\
&& ~~~~~~~~~~~~~~~~- e^{i (\al \cos(\om t + \pi/3) + (k_x -\sqrt{3}
k_y)/2)} \non \\
&& ~~~~~~~~~~~~~~~~- e^{-i(\al \cos(\om t) + k_x)}, \label{graham1} \eea
where $\tau_\pm = (1/2) (\tau_x \pm i \tau_y)$. Defining $B_\vk (t) = {\rm
Re} [Z_\vk (t)]$ and $C_\vk (t) = {\rm Im} [ Z_\vk (t)]$, we obtain
\beq H_\vk (t) ~=~ B_\vk (t) ~\tau_x ~-~ C_\vk (t) \tau_y. \eeq We
can perform a unitary transformation to convert this to \beq H_\vk
(t) ~=~ -~ C_\vk (t) \tau_z ~+~ B_\vk (t) ~\tau_x. \label{hamvk}
\eeq We will use the form in Eq.~\eqref{hamvk} in the rest of this
section; the fact that $\tau_x$ and $\tau_z$ are real and symmetric
will prove to be very useful. Note that the instantaneous
eigenvalues of the Hamiltonian in Eq.~\eqref{hamvk} are given by
$E_{\vec k}^{\pm}= \pm ~| Z_\vk (t)|$. We note here that both the
diagonal and off-diagonal terms in Eq.\ \eqref{hamvk} depends
explicitly on time; thus irradiated graphene constitutes an example
of the periodic version of the two-rate protocol studied in context
of other models in Ref.\ \onlinecite{sau1}.

In the rest of this section we will discuss the symmetries of the
momentum space Hamiltonian $H_\vk (t)$ and the corresponding time evolution
operator
\beq U_\vk (t,0) ~\equiv~ U_{\vec k}(t) = {\cal T}_t ~\exp [- i \int_0^t dt'
H_\vk (t')] \label{ut1}, \eeq
at various points $\vk = (k_x,k_y)$ in the Brillouin zone. Note
that under $\vk \to \vk + {\vec G}_1$ or $\vk +{\vec G}_2$, $Z_\vk
(t) \to e^{-i2\pi/3} Z_\vk (t)$. Hence the symmetries of $H_\vk (t)$
are similar for points in $\vk$ space which are related to each
other by reciprocal lattice vectors. When looking for symmetries, we
will consider arbitrary values of $t$ in Eq.~\eqref{ut1}, not just
$t=T$. Further, we will look for symmetries which hold for all
values of $\al$ and $T$.

{\it $\Ga$ point}: The Hamiltonian $H_{\vec k}$ at the $\Ga$
point $[(k_x,k_y)=(0,0)]$ has several underlying symmetries which
can be deduced by an inspection of $H_{\vec k}$ and $Z_{\vec k}$ (Eq.\
\eqref{graham1}). These are given by
\bea H_\vk (T-t) &=& H_\vk (t) \, \, [{\rm since} \, Z_\vk (T-t) = Z_\vk
(t)] \label{symgam1} \\
H_\vk (T/6-t) &=& H_\vk (T/6+t)= \tau_x H_\vk (t) \tau_x
\non \\
&& [{\rm since}\, Z_\vk (T/6- t) = Z_\vk^* (t)]. \label{symgam2} \\
H_\vk (T/3-t) &=& H_\vk (T/3+t)= H_\vk (t). \label{symgam3} \\
H_\vk (T/2-t) &=& H_\vk (T/2+t)= \tau_x H_\vk (t) \tau_x.
\label{symgam4}\\
H_\vk (2T/3-t) &=& H_\vk (2T/3+t)= H_\vk (t). \label{symgam5} \\
H_\vk (5T/6-t) &=& H_\vk (5T/6+t) = \tau_x H_\vk (t) \tau_x.
\label{symgam6} \eea The above relations imply that the
instantaneous eigenvalues of $H_\vk (t)$ are identical at times
$nT/6 \pm t$ for $n=1, 2, \cdots, 5$, and at $T - t$. These
relations also lead to the following symmetry properties of $U(t)$.
On the right hand side of Eq.~\eqref{ut1}, we can divide the
integral in the exponential into $N_t$ steps, each of size $\De t =
t/N_t$. Defining $t_j = (j-1/2)\De t$, we can write Eq.~\eqref{ut1}
as \bea U(t) &=& e^{-i \De t H(t_{N_t})} ~e^{-i \De t H(t_{N_t-1})}
~\cdots
\non \\
&& \cdots ~e^{-i \De t H(t_2)} ~e^{-i \De t H(t_1)} \label{ut2} \eea
in the limit $N_t \to \infty$. Eq.\ \eqref{symgam1} and the symmetry
of the Pauli matrices in Eq.~\eqref{hamvk} implies that \beq
H(t_{N_t +1 -j}) ~=~ H(t_j) ~=~ [H(t_j)]^T, \eeq where the
superscript $T$ means transpose. This, in turn, indicates that \beq
[U(T)]^T ~=~ U(T), \eeq which means that $U(T)$ is of the form \beq
U(T) ~=~ \pm \exp [i(d_1 \tau_x + d_3 \tau_z)], \label{ut3} \eeq
namely, $\tau_y$ does not appear in the exponential. In
Eq.~\eqref{ut3}, the parameters $d_1, ~d_3$ are real and satisfy $0
\le \sqrt{d_1^2 + d_3^2} < \pi$. We have allowed for a $\pm$ sign in
Eq.~\eqref{ut3} to ensure that $\sqrt{d_1^2 + d_3^2}$ is strictly
less than $\pi$.

Since only two parameters, $d_1$ and $d_3$, appear in
Eq.~\eqref{ut3}, there is a possibility that we can make both
parameters equal to zero by suitably choosing the two parameters
$\al$ and $T$ appearing in the definition of $U(T)$ in
Eq.~\eqref{ut1}. This means that there is a possibility of varying
$\al$ and $T$ to make $U(T)= \pm I$ ($I$ denotes the $2 \times 2$
identity matrix), so that there is a phase band crossing at $t=T$.

Similar arguments based on Eqs.\ \eqref{symgam3} and \eqref{symgam4}
imply that $U(T/3)$ and $U(2T/3)$ also have the two-parameter form
in Eq.~\eqref{ut3} and can therefore be made equal to $\pm I$ by
choosing $\al$ and $T$ appropriately. From Eqs.\ \eqref{symgam3} and
\eqref{symgam5}, we also find that \bea U(2T/3) &=& [U(T/3)]^2 \quad
U(T) = [U(T/3)]^3. \label{ugammacond1} \eea Hence, $U(T/3) = \pm I$
implies that $U(T) = \pm I$; hence a phase band crossing at $t=T/3$
implies a crossing at $t=T$. The converse is not necessarily true;
we can have $U(T) = \pm I$ without having $U(T/3) = \pm I$.

Next, we can use Eq.\ \eqref{symgam4} and the expression in
Eq.~\eqref{ut2} to show that \beq U(T/2) ~=~ \tau_x [U(T/2)]^T
\tau_x. \eeq This implies that $U(T/2)$ has the form \beq U(T/2) ~=~
\pm \exp [i(d_1 \tau_x + d_2 \tau_y)], \label{ut4} \eeq namely,
$\tau_z$ does not appear in the exponential. Thus $U(T/2)$ also has
a two-parameter form; hence it may be possible to find $\al$ and $T$
so that $U(T/2) = \pm I$ and thereby have a phase band crossing at
$t=T/2$. Eq.\ \ref{symgam4} also implies that \beq U(T) ~=~ \tau_x
U(T/2) \tau_x U(T/2). \label{ut5} \eeq Eqs.~\eqref{ut4} and
\eqref{ut5} imply that if we can make $d_1 = 0$, we will have $U(T)
= I$. Thus we only need to make one parameter, $d_1$, equal to zero
by varying $\al$ and $T$ in order to make $U(T) = I$ and so have a
phase band crossing at $t=T$. Hence we expect that there may be a
line in the $\al-T$ plane where there is a phase band crossing at
$t=T$. Similar arguments based Eqs.\ \ref{symgam2} and \ref{symgam6}
imply that $U(T/6)$ and $U(5T/6)$ also have the two-parameter form
in Eq.~\eqref{ut4} and can therefore be made equal to $\pm I$ by
choosing $\al$ and $T$ appropriately.

A key point that emerges from the discussion above is that whenever
we find a fraction $f$ (lying in the range $0 < f \le 1$) such that
either $Z_\vk (fT-t) = Z_\vk (t)$ (implying $H_\vk (fT-t) = H_\vk
(t)$) or $Z_\vk (fT-t) = Z_\vk^* (t)$ (implying $H_\vk (fT-t) =
\tau_x H_\vk (t) \tau_x$), $U(fT)$ will have a two-parameter form
given by either Eq.~\eqref{ut3} or \eqref{ut4}. One then expects
that there would be points in the $\al-T$ plane where $U(fT)= \pm I$
so that there are phase band crossings at $t=fT$. However, we
note that such a two-parameter form of $U(fT)$ is not sufficient to
have a phase band crossing; we shall discuss this point in detail in
Sec.\ \ref{secadimp}.

{\it Dirac ($K$ and $K'$) points}: These are the two inequivalent
points in the Brillouin zone where the conduction and valence bands
of graphene touch in the absence of any external radiation; their
positions in the Brillouin zone are given by $\vk = (0, \pm
4\pi/(3\sqrt{3}))$. Considering the point $K = (0,
4\pi/(3\sqrt{3}))$, we find the following symmetries.
\begin{widetext}
\bea H_\vk (T/2-t) &=& \tau_x H_\vk (t) \tau_x \label{symdir1} \\
H_\vk (T/6-t) &=& [(1/2) \tau_x - (\sqrt{3}/2) \tau_z] H_\vk (t)
[(1/2)\tau_x - (\sqrt{3}/2) \tau_z] \quad [{\rm since}\, Z_\vk
(T/6-t) = e^{i2\pi/3} Z_\vk^* (t)]. \label{symdir2}\\
H_\vk (T/3+t) &=& e^{(i\pi/3) \tau_y} H_\vk (t) e^{-(i\pi/3)
\tau_y} \quad [{\rm since} Z_\vk (T/3+t) = e^{-i2\pi/3} Z_\vk (t)]
\label{symdir3} \\
H_\vk (2T/3+t) &=& e^{-(i\pi/3) \tau_y} H_\vk (t) e^{i(\pi/3)
\tau_y} \quad [{\rm since}\, Z_\vk (2T/3+t) = e^{i2\pi/3} Z_\vk (t)]
\label{symdir4} \\
H_\vk (5T/6-t) &=& [(1/2) \tau_x + (\sqrt{3}/2) \tau_z] H_\vk (t)
[(1/2) \tau_x - (\sqrt{3}/2) \tau_z] \quad [{\rm since} \, Z_\vk
(5T/6-t) = e^{-i2\pi/3} Z_\vk^* (t).] \label{symdir5} \eea
\end{widetext}
Hence the eigenvalues of $H(t)$ are identical at the times $T/6-t$,
$T/3+t$, $T/2-t$, and $5T/6-t$. However, from the point of view of
phase band crossings, the only useful relation is Eq.\
\eqref{symdir1}. This implies that $U(T/2)$ is of the form given in
Eq.~\eqref{ut4}, and we can have a phase band crossing at $t=T/2$
(apart from those at $t=T$) at certain points in the $\al-T$ plane.
The symmetries of the $K'$ points are identical.

{\it $M$ points}: There are 3 sets of inequivalent points, namely
$M_{1,2,3}$. Out of these, the points $M_3$ lie at $\vk=(\pm 2\pi/3,
0)$. To be specific, we consider the point at $\vk = (2\pi/3,0)$. We
find the following symmetries.
\bea && H_\vk (T-t) = H_\vk (t) \label{symcor1} \\
&& H_\vk (T/2-t) = H_\vk (T/2+t) =[(1/2) \tau_x - (\sqrt{3}/2) \tau_z] \non \\
&& \times H_\vk (t) [(1/2) \tau_x - (\sqrt{3}/2) \tau_z]. \label{symcor2} \eea
Hence the eigenvalues of $H(t)$ are identical at $T-t$ and $T/2 \pm
t$. Eq.\ \eqref{symcor1} implies that $U(T)$ is of the form given in
Eq.~\eqref{ut3}, and we can have a phase band crossing at $t=T$ at
certain points in the $\al-T$ plane.

By rotating the $M_3$ points by $\pm 2\pi/3$, we obtain four other
points; these are related pairwise by reciprocal lattice vectors.
Hence we only have to consider two points, say, $\vk = (\pi/3,\pm
\pi/ \sqrt{3})$. These are the $M_{1,2}$ points alluded to in the
last paragraph. We find the following symmetries at $(\pi/3,\pi/
\sqrt{3})$.
\bea && H_\vk (T/3-t) = H_\vk (t) \label{symmid1} \\
&& H_\vk (T/2+t) = H_\vk (5T/6-t) = [(1/2) \tau_x + (\sqrt{3}/2) \tau_z] \non \\
&& \times H_\vk (t) [(1/2) \tau_x - (\sqrt{3}/2) \tau_z]. \label{symmid2} \eea
Hence the eigenvalues of $H(t)$ are identical at the times $T/3-t$,
$T/2+t$, and $5T/6-t$. Further, $U(T/3)$ has the form given in
Eqs.~\eqref{ut3}, implying that there can be phase band crossings at
$t=T/3$ at certain points in the $\al-T$ plane.

{\it $X$ point}: As shown in Fig.\ \ref{fig1}, this point
corresponds to the midpoint of the line joining the $\Ga$ point and
the Dirac point lying at $(k_x,k_y)= (2\pi/3, 2 \pi/(3 \sqrt{3}))$;
it has coordinates given by $(k_x,k_y)=(\pi/3, \pi/(3 \sqrt{3}))$.
At this point we find that $Z_{\vec k}(T/6-t)= Z^{\ast}_{\vec k}
(t)$ which leads to \bea H_{\vec k}(T/6-t) &=& H_{\vec k} (7T/6-t) =
\sigma_x H_{\vec k} (t) \sigma_x. \label{symx1} \eea We can
therefore expect phase band crossings at $t=T/6$ and $t=7T/6$ at
some suitably chosen points in the $\al-T$ plane.

{\it Line given by $\vk = (k_x,0)$}: We find only the symmetry $Z_\vk (T-t)=
Z_\vk (t)$ and $H_\vk (T-t)=H_\vk (t)$ for an arbitrary point on the line
$k_y = 0$ (but not at the $\Ga$ point or the $M$ points
which have a larger symmetry as we have seen above). Hence the eigenvalues of
$H(t)$ and $H(T-t)$ are identical, and $U(T)$ has the form given in
Eq.~\eqref{ut3}.

{\it Line given by $\vk = (0,k_y)$}: We find only the symmetry $Z_\vk (T/2-t)=
Z_\vk^{\ast} (t)$ and $H_\vk (T/2-t)= \tau_x H_\vk (t) \tau_x$ for an
arbitrary point on the line $k_x = 0$ (but not at the Dirac points which
have a larger symmetry as discussed above). The eigenvalues of $H(t)$ and
$H(T/2-t)$ are therefore identical, and $U(T/2)$ has the form given in
Eq.~\eqref{ut4}.

{\it Line given by $\vk = (k_x,\sqrt{3}k_x)$}: We find only the symmetry
$Z_\vk (T/3-t)= Z_\vk (t)$ and $H_\vk (T/3-t)=H_\vk (t)$ for an arbitrary
point on the line $k_y = \sqrt{3} k_x$ (but not at the points $\vk =
\pm (\pi/3, \pi/\sqrt{3})$ which have a larger symmetry as discussed
above). Hence the eigenvalues of $H(t)$ and $H(T/3-t)$ are identical, and
$U(T/3)$ has the form given in Eq.~\eqref{ut3}.

Finally, in all the cases where at some value of $\vk$, $U_\vk (T)$
has the form given in either Eq.~\eqref{ut3} or Eq.~\eqref{ut4} and
we have a phase band crossing at $t=T$ at a certain point in the
$\al-T$ plane, we can find two more points in the $\vk$ space where
there will be a phase band crossing at the same values of $\al$ and
$T$. This is because $Z_\vk (t)$ in Eq.~\eqref{graham1} is invariant
under a simultaneous rotation by $2\pi/3$ in $\vk$ space and a shift
in $t$ by $T/3$. More precisely, we find that
\beq Z (k_x,k_y,t) = Z (-\frac{k_x}{2}+\frac{\sqrt{3}k_y}{2},-\frac{k_y}{2}-
\frac{\sqrt{3}k_x}{2}, t-\frac{T}{3}). \eeq
Next, we note that if we hold $\vk$ fixed, set $t=T$ and shift $t' \to
t' + s$ in Eq.~\eqref{ut1}, thus defining
\beq U_\vk (T; s) ~\equiv~ {\cal T} ~\exp [- i \int_s^{T+s} dt' H_\vk (t')],
\label{ut6} \eeq
then $U_\vk (T)$ in Eq.~\eqref{ut1} and $U_\vk (T;s)$ in Eq.~\eqref{ut6}
have the same eigenvalues. (Their eigenvectors are related by a
unitary transformation by the operator ${\cal T} \exp (-i \int_0^s
dt' H_\vk (t')])$). This means that if $U(k_x,k_y,T) = \pm I$ with
eigenvalues $\pm 1$ for some value of $\al$ and $T$, then
$U(-k_x/2+\sqrt{3}k_y/2,-k_y/2 -\sqrt{3}k_x/2,T)$ will also have the
eigenvalues $\pm 1$. Thus a phase band crossing at $t=T$ at some value of
$\vk$ means that there is also a phase band crossing at $t=T$ at values of
$\vk$ obtained by $\pm 2\pi/3$ rotations. (This is ultimately related to the
fact that the graphene lattice is invariant under rotations by $\pm 2\pi/3$).
Thus, for example, the discussion of the line given by $\vk = (k_x,0)$ show
that we can also have phase band crossings at $t=T$ on the
two lines obtained by rotating the line $\vk = (k_x,0)$ by $\pm 2\pi/3$.

In the next section we use these symmetry properties to understand the
condition of crossing of the phase bands of $U_{\vec k} (t)$ at different
symmetry points.

\section{Adiabatic-Impulse Method}
\label{secadimp}

In this section, we develop the adiabatic-impulse method and use it
to compute the phase bands for a Dirac Hamiltonian in the presence
of radiation in the low-frequency regime. The advantage of this
method lies in the fact that it becomes accurate in the
low-frequency regime where standard $1/\om$ expansions fail; thus
this method serves as a complimentary analytic tool for
understanding the low-frequency response of periodically driven
closed integrable quantum systems. This method is known to be
accurate for $\al T, T \ge 2 \pi$ (in units of $\hbar/\ga$)
\cite{nori1} and has already been applied for phase band
computations in periodically driven systems where a single parameter
of the Hamiltonian (usually present in the diagonal term of $H_{\vec
k}$) is varied as a function of time \cite{bm1}. However, as can be
seen from Eq.\ \eqref{hamvk}, the Hamiltonian of graphene in the
presence of external radiation has both diagonal and off-diagonal
terms varying as functions of time; thus it cannot be mapped to the
class of driven Hamiltonians studied in Ref.\ \onlinecite{bm1}. In
particular, the application of the adiabatic-impulse approximation
to such systems requires a separate analysis which we now chart out.

The adiabatic-impulse approximation \cite{nori1} proceeds by
estimating the gap between the instantaneous energy eigenvalues of
the driven system. The instantaneous energy eigenvalues $E_{\vec
k}^{\pm} (t) = \pm E_{\vec k} (t)$ and eigenvectors $ \psi_{\vec
k}^{\pm} (t) = (u_{\vec k}^{\pm} (t), v_{\vec k}^{\pm} (t))^T$ of
$H'_{\vec k} (t)$ are given by (Eq.\ \eqref{hamvk})
\bea E_{\vec k} (t) &=& \sqrt{B_{\vec k}^2 (t) + C_{\vec k}^2 (t)}, \non \\
u^-_{\vec k} (t) &=& - C_{\vec k} (t)/\sqrt{(E_{\vec k} (t) +B_{\vec k} (t))^2
+ C_{\vec k} (t)^2}, \non \\
v^-_{\vec k} (t) &=& (E_{\vec k} (t)+ B_{\vec k} (t))/\sqrt{(E_{\vec k} (t) +
B_{\vec k} (t))^2+ C_{\vec k} (t)^2}, \non \\
u_{\vec k}^+ (t) &=& -v_{\vec k}^- (t), \quad v_{\vec k}^+ (t) =
u_{\vec k}^- (t). \label{eigen1} \eea The eigenvectors $\psi_k^{\pm}
(t)$ can be used to construct the adiabatic basis. The wave
function, $\Psi_{\vec k} (t)$ of the driven system can be written in
this basis as \bea \Psi_{\vec k} (t) = c_{1 \vec k} (t) \psi_{\vec
k}^- (t) + c_{2 \vec k} (t) \psi_{\vec k}^+ (t), \label{adrep1}
\eea where $c_{1 (2) \vec k} (t)$ represents the overlap of the wave
function with the instantaneous ground (excited) states. The initial
condition, for a system starting in the local ground state at $t=0$,
is thus $ c_{1\vec k} (0)=1$ and $c_{2 \vec k} (0)=0$. The time
evolution of the wave function $\Psi_{\vec k} (t)$ in the adiabatic
basis can be understood by tracking the time dependence of the
coefficients $c_{1 (2)\vec k} (t)$. As shown in Refs.\
\onlinecite{bm1} and \onlinecite{nori1}, we can relate $\vec c_{\vec
k} (t) = (c_{1\vec k} (t), c_{2 \vec k} (t))^T$ to $\vec c_{\vec k}
(0)$ through a evolution matrix $U^{\rm ad}_{\vec k} (t)$ given by
\bea \vec c_{\vec k} (t) = U_{\vec k}^{\rm ad}(t,0) \vec c_{\vec k}
(0). \eea Note that $U_{\rm k}^{\rm ad}(t,0)$ is related to the
evolution operator $U_{\vec k }(t,0)$ defined as \bea \Psi_{\vec k}
(t) = U_{\vec k}(t,0) \Psi_{\vec k} (0)\eea through the overlap
$\eta_{\vec k} (t)$ of the ground states in the adiabatic and
diabatic bases as \cite{bm1} \bea U_{\vec k}(t,0) &=& \left[
\eta_{\vec k} (t) I - i \tau_y \sqrt{1-\eta_{
\vec k}^2 (t)}\right] U_{\vec k}^{\rm ad}(t,0), \non \\
\eta_{\vec k} (t) &=& (\psi_{\vec k}^{- \ast} (t))^T \psi_{\vec k}^- (0)
\non \\
&=& u^{\ast -}_{\vec k} (t) u^-_{\vec k} (0) + v^{\ast -}_{\vec k} (t)
v^-_{\vec k} (0), \label{etadef} \eea
where $\tau_{x,y,z}$ denote Pauli matrices and $I$ is the $2 \times 2$
identity matrix.

We now envisage a situation where the system goes through $n$
avoided level crossings during a drive cycle. These avoided level
crossings allow us to divide the time evolution of the system within
a drive period into $n+1$ regions which separate these crossings.
The key assumption of the adiabatic-impulse approximation is that in
these $n+1$ regions, the systems undergoes adiabatic evolution and
no excitations are produced \cite{nori1}. The evolution operator in
the $m^{\rm th}$ region can thus be written in the adiabatic basis
as \cite{nori1} \bea U_{\vec k}^{(m) \rm ad}(t,t_0) &=& \exp[-
i\tau_z \xi^{(m)}_{\vec k}
(t,t_0)], \non \\
\xi^{(m)}_{\vec k}(t,t_0) &=& \int_{t_0}^t dt' E_{\vec k}(t').
\label{uad1} \eea
Note that $\xi^{(m)}_{\vec k}(t,t_0)$ denotes the kinematic phase
picked up by the wave function during the evolution, and the superscript
$m$ indicates that both $t$ and $t_0$ lie in the $m^{\rm th}$
adiabatic regime. For later use, we also define $\xi^{(i)}_{\vec
k}(t_{i \vec k},t_{i-1 \vec k}) \equiv \xi_{i \vec k}$ and
$\xi^{(i)}_{\vec k}(t,t_{i-1 \vec k}) \equiv \xi_{i \vec k} (t)$.

At the avoided level crossing points separating two adiabatic
regions, the evolution becomes non-adiabatic. The approximation
involved in the present method is that it treats the impulse regions
as isolated points where the avoided level crossings occur. Clearly,
such an approximation becomes better at lower frequencies, and hence
this method works well at small $\om$ \cite{nori1}. The excitation
probability at these points is usually estimated by linearizing the
drive term around these regions which allows for an analytic
calculation of the excitation probability (when a single (diagonal)
term of the system Hamiltonian varies with time) via the
Kibble-Zurek approach \cite{nori1,bm1}. However, the present class
of systems, where both diagonal and off-diagonal elements of the
Hamiltonian vary with time, calls for a modification of this
procedure. To demonstrate this modification, we consider the
Hamiltonian given by Eq.\ \eqref{hamvk} near the $j^{\rm th}$
avoided level crossing which separates the $(j-1)^{\rm th}$ and
$j^{\rm th}$ adiabatic regions. The time $t_{j \vec k}$ at which
such a crossing occurs can be obtained by setting $dE_{\vec
k}(t)/dt=0$; this leads to (Eq.\ \eqref{eigen1}) \bea C_{\vec
k}(t_{j \vec k}) \, \frac{d C_{\vec k}(t_{j \vec k})}{dt} + B_{\vec
k}(t_{j \vec k}) \, \frac{d B_{\vec k}(t_{j \vec k})}{dt} &=& 0.
\label{lcross} \eea We now follow Ref.\ \onlinecite{nori1} to
linearize $H_{\vec k} (t)$ around $t=t_{j \vec k}$. This leads to an
effective Hamiltonian $H^{\rm eff}_{\vec k} (t)$ given by \bea
H^{\rm eff}_{\vec k} (t) &=& [C_{\vec k}(t_{j \vec k}) + (t-t_{j
\vec
k}) \frac{d C_{\vec k}(t_{j \vec k})}{dt} ] \tau_z \non \\
&& + ~[B_{\vec k}(t_{j \vec k}) + (t-t_{j \vec k}) \frac{d B_{\vec
k}(t_{j \vec k})}{dt} ] \tau_x. \label{effham} \eea Note that
$H^{\rm eff}_{\vec k} (t)$, which determines the excitation
probability near $t_{j \vec k}$, is not of the Kibble-Zurek form in
the sense that both its diagonal and off-diagonal terms depend
explicitly on time. To cast it into this form, we define a new set
of Pauli matrices $\vec \sigma$; using these matrices we can write
$H^{\rm eff}_{\vec k} (t)$ as \bea H^{\rm eff}_{\vec k} (t) &=&
\nu_{1\vec k} (t-t_{j \vec k}) \sigma_z + \nu_{2 \vec k} \sigma_x,
\label{sigmaham} \eea where $\nu_{1 (2) \vec k}$ are independent of
time \cite{sau1}. A comparison between Eqs.\ \eqref{effham} and
\eqref{sigmaham} shows that \bea \sigma_z \nu_{1 \vec k} &=& \tau_z
\frac{dC_{1 \vec k}(t_{j \vec k})}{dt}
+ \tau_x \frac{dB_{1 \vec k}(t_{j \vec k})}{dt}, \non \\
\sigma_x \nu_{2 \vec k} &=& \tau_z C_{\vec k}(t_{j \vec k}) + \tau_x
B_{\vec k}(t_{j \vec k}). \label{sigrel} \eea
Using the identities ${\rm Det} \sigma_3 = {\rm Det} \sigma_x=-1$,
we can then determine
\bea \nu_{1 \vec k} &=& [( \frac{dC_{1 \vec k}(t_{j \vec k})}{dt} )^2 ~+~
( \frac{dB_{1 \vec k}(t_{j \vec k})}{dt} )^2]^{1/2}, \non \\
\nu_{2 \vec k} &=& [B^2_{\vec k}(t_{j \vec k}) + C^2_{\vec k}(t_{j
\vec k})]^{1/2}. \label{nudef} \eea
Note that the anticommutation relation $\{\sigma_z, \sigma_x\}=0$ is
satisfied due to Eq.\ \eqref{lcross}. We can then read off the
Landau-Zener excitation probability from Eq.\ \eqref{sigmaham},
\bea p_{j\vec k} &=& e^{-2 \pi r_{j\vec k}}, \quad r_{j\vec k}=
\nu^2_{2 \vec k}/(2 \nu_{1\vec k}). \label{lzprob} \eea
Having obtained these probabilities, we follow Refs.\ \onlinecite{bm1}
and \onlinecite{nori1} to construct a transfer matrix $S_{j\vec
k}$ which relates $\vec c_{\vec k}(t_{j \vec k} -\epsilon)
\equiv \vec c^{\,\,j}_{\vec k}$ to $\vec c_{\vec k}(t_{j \vec k} +
\epsilon) \equiv \vec c^{\,\,j+1}_{\vec k}$,\cite{nori1,bm1}
\bea \vec c^{\,\,j+1}_{\vec k} &=& S_{j\vec k} \vec c^{\,\,j}_{\vec k},
\label{crel1}\\
S_{j\vec k} &=& I \sqrt{1-p_{j\vec k}} e^{-i \tau_z \Phi_{j\vec k}}
- i
\tau_y \sqrt{p_{j\vec k}}, \non \\
\Phi_{j\vec k} &=& r_{j\vec k} ( 1 - \ln r_{j\vec k}) + {\rm Arg}(1
- i\Ga[r_{j\vec k}]) - 3 \pi/4. \non \eea Here $\Phi_{j\vec k}$ is
the Stuckelberg phase \cite{stu1,nori1} generated at the $j^{\rm
th}$ avoided crossing, and $S_{j\vec k}$ can be viewed as the
transfer matrix which takes the wave function across such a
crossing.

Combining Eqs.\ \eqref{crel1} and \eqref{uad1}, the coefficients $\vec
c_{\vec k}$ after $n$ crossings at $t=t_f$ are found to be
\bea \vec c_{\vec k}(t_f) &=& U^{(n+1) \rm ad}_{\vec k}(t_f, t_n(\vec
k)) (S_{n\vec k})^T ~\cdots \non \\
&& \times ~S_{1\vec k}\, U^{(1) \rm ad}_{\vec k}(t_1 (\vec k),0)
\left( \begin{array}{c} 1 \\ 0 \end{array} \right) \non \\
&=& U_{\vec k}^{\rm ad}(t_f,0) \left( \begin{array}{c} 1 \\ 0
\end{array} \right), \label{uadrel1} \eea
where $U_{\vec k}^{\rm ad}(t_f,0)$ is the evolution operator in the
adiabatic basis, and $S^T$ denotes the transpose of $S$. Using Eqs.\
\eqref{etadef} and \eqref{uadrel1}, we may obtain the evolution
operator for the system at $t=t_f$ in terms of $r_{\vec k}$ (in
Eqs.\ \eqref{lzprob} and \eqref{crel1}) and $\xi_{\vec k}(t_1,t_2)$
(in Eq.\ \eqref{uad1}). The eigenvalues of $U_{\vec k}(t_f,0)$ can be
obtained by diagonalizing the $2 \times 2$ matrix obtained. The
unitarity of $U_{\vec k}(t_f,0)$ ensures that these eigenvalues or
phase bands are given by
\bea \lambda_{\pm}(\vec k,t) &=& \exp[\pm i \phi(\vec k,t)],
\label{phasebands1} \\
\cos (\phi(\vec k,t)) &=& {\rm Re}[U_{\vec k}(t,0)]_{11}) \non \\
&=& \eta_{\vec k}(t_f) c_{1 \vec k}(t_f) + \sqrt{1-\eta^2_{\vec
k}(t_f)} c_{2 \vec k}(t_f). \non \eea
The details of the computation
of the phase bands using Eqs.\ \eqref{etadef}, \eqref{crel1}, and
\eqref{uadrel1} are charted out in the Appendix. Here we present the
analytic expression for the case $n=2$. For $t_{2\vec k} \le t_f \le
T=2\pi/\om$, the expression for the phase bands are given by
\begin{widetext}
\bea &&\cos(\phi^{(2)}(\vec k,t_f)) ~=~ \eta_{\vec k}(t_f) \left( \sqrt{
(1-p_{1\vk})(1-p_{2\vk})} ~\cos[\Phi^s_{\vec k}+ \xi^s_{\vec k}(t_f)] ~+~
\sqrt{p_{1\vk}p_{2 \vec k}} \cos[\xi^s_{\vec k}(t_f) - 2 \xi_{2 \vec k}]
\right) \label{pbn2exp} \\
&& + \sqrt{1-\eta_{\vk}^2 (t)} \left( \sqrt{p_{1\vk}(1-p_{2\vk})} ~\cos(
\xi_{2 \vec k}+ \xi_{3 \vec k}(t_f)-\xi_{1\vec k} -\Phi_{2 \vec k}) ~-~
\sqrt{p_{2\vk}(1-p_{1\vk})} ~\cos[ \Phi_{1 \vec k} + \xi_{1 \vec k} +
\xi_{2 \vec k} -\xi_{3 \vec k}(t_f)] \right), \non \eea
\end{widetext}
where $\Phi^s_{\vec k} = \sum_{i=1, 2} \Phi_{i \vec k}$, and
$\xi_{\vec k}^s(t_f) = \sum_{i=1, 2} \xi_{i \vec k} + \xi_{3 \vec
k}(t_f)$. We shall analyze Eq.\ \eqref{pbn2exp} (and its
counterparts for different $n$) to obtain general phase band
crossing conditions for irradiated graphene in Sec.\ \ref{secphd}.

To find the conditions for phase band crossings, we note that
Eq.\ \eqref{phasebands1} indicates that the condition for
$\cos[\phi^{(n)}(\vec k, t)]=\pm 1$ can be understood by finding its
maximum/minimum values; these occur when
\bea \eta_m = \pm {\rm Re}[c_{1\vec k}(t_f)]/\sqrt{({\rm Re}[c_{1\vec
k}(t_f)])^2 + ({\rm Re}[c_{2\vec k}(t_f)])^2}. \non \\
\eea
The corresponding extremum value of $\cos[\phi^{(n)}(\vec k, t_f)]$ is given by
\bea \cos[\phi^{(n) e}(\vec k,t_f)]= \pm \sqrt{({\rm Re}[c_{1\vec
k}(t_f)])^2 + ({\rm Re}[c_{2\vec k}(t_f)])^2}. \non \\
\eea Note that since $|c_{1 \vec k} (t)|^2+|c_{2 \vec k} (t)|^2=1$,
the condition $\cos[\phi^{(n) e}(\vec k,t_f)]=\pm 1$ thus requires
$c_{1 \vec k} (t)$ and $c_{2 \vec k} (t)$ to be real (apart from a
possible irrelevant global phase). For a generic time evolving wave
function, this is most likely when one of its components vanish.
This leaves us with two possibilities. The first is $\eta_{\vec
k}(t_f) =0= {\rm Re}[c_{1\vec k}(t_f)]$ and the second is
$\eta_{\vec k}(t_f) = 1= \pm {\rm Re}[c_{1\vec k}(t_f)]$. The former
possibility requires that the excited state of $H_{\vec k}(t_f)$ be
the same as the ground state of $H_{\vec k} (0)$, and this is not
guaranteed by any symmetry of the Hamiltonian unless $p_{\vec k}=1$
at some crossing point. The latter condition, in contrast, is
generic at $t_f=T$ for any $\vec k$ and at $t=nT/3$ for any integer
$n$ at the $\Ga$ point. As we shall see numerically, this latter
condition is always satisfied at all the phase band crossings that
we find. In what follows, we thus find the generic expression for
the phase bands after $n$ avoided level crossings when $\eta_{\vec
k}(t_f)= 1$. The detailed method of doing so is sketched in the
Appendix. The final result that we obtain from this procedure is as
follows. For an even number ($2n$) of avoided level crossings,
denoting $\phi^{(2n)}(\vec k, t_f) \equiv \phi^e$, we get
\begin{widetext}
\bea \cos (\phi^e) &=& \sum_{jmax=0, 2, \cdots, 2n} \sum_{\al}
\prod_{j_{\al} = 1}^{jmax} (1-p_{j_{\al} \vec k})^{1/2}
\prod_{j'_{\al} \ne j_{\al}=1}^{2n-jmax} p_{j'_{\al} \vec
k}^{1/2} (-1)^{n_1} \cos\left[\Phi_{\vec k}^s + \xi_{\vec k}^s(t_f)
-\sum_a (\ga_a \Phi_{a \vec k} + \de_a \xi_{a \vec k}(t_f))
\right], \non \\
\label{evencos1} \eea
\end{widetext}
where the sum over the index $\al$ represents a sum over all
possible permutations of $j_{\al}$ and $j'_{\al}$ for a fixed
$jmax$, and
\bea \Phi^s_{\vec k} &=& \sum_{i=1, 2n} \Phi_{i \vec k}, \quad
\xi_{\vec k}^s(t_f) = \sum_{i=1, 2n} \xi_{i \vec k} + \xi_{2n+1
\vec k}(t_f), \non \\
\phi^s_{\vec k} &=& \sum_{i=1, 2n} \phi_{i \vec k}, \quad \xi_{\vec
k}^s = \sum_{i=1, 2n+1} \xi_{i \vec k}. \label{phixieq1} \eea
In Eq.\ \eqref{evencos1}, $n_1 = {\rm Max}[j'_{\al}]-{\rm
Min}[j'_{\al}]+1$ provided ${\rm Min}[j'_{\al}] \ne 0$ and is $0$
otherwise, and the coefficients $\ga_a$ and $\de_a$ for any
given permutation $\al$ are given by
\bea \ga_a &=& 1 \, \, {\rm for}\, a \in j'_{\al}, \non \\
&=& 2 \, \, {\rm for}\, a \in j_{\al} \,{\rm with}\, j^{o \prime}_{\al} <
j_{\al} < j^{e \prime}_{\al}, \label{evencoeff} \\
&=& 0 \, \, {\rm otherwise}, \non \\
\de_a &=& 2 \, \, {\rm for}\, a \in j_{\al},j'_{\al}\,\,
{\rm with} \, {\rm
Min}[j'_{\al}] < j_{\al},j'_{\al} \le {\rm Max}[j'_{\al}] \non \\
&=& 0 \,\, {\rm if}\, a \in j_{\al},j'_{\al}\, {\rm with}\,
a-1 \in j'_{\al}\, {\rm and}\, \de_{a-1}^e =2, \non
\eea
where $j^{o \prime}_{\al}$ denotes any odd occurrence of $j'$
during a permutation, and $j^{e \prime}_{\al}$ denotes its next
occurrence in that permutation. Note that for $n=1$, Eq.\
\eqref{evencoeff} reproduces Eq.\ \eqref{pbn2exp} for $\eta_{\vec k}(t_f)=1$.

\begin{figure}[t!]
\begin{center}
\includegraphics[width=\columnwidth]{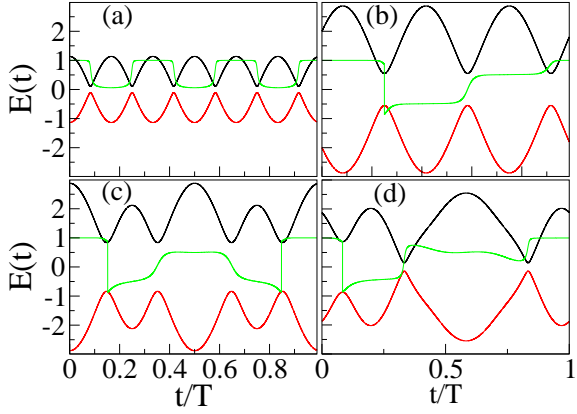}
\end{center}
\caption{Plot of instantaneous ground (red solid
line) and excited (black solid line) state energies and the wave
function overlap $\eta_{\vec k} (t)$ (green dashed line) as a
function of $t/T$ for (a) $\Ga$ point with $(k_x, k_y)=(0,0)$ for
the dimensionless radiation amplitude $\al=2.35$, (b) Dirac ($K$)
point with $(k_x,k_y)=(0,4 \pi/3 \sqrt{3})$ for $\al=2$, (c) $M$ point
with $(k_x,k_y)=(2 \pi/3 , 0)$ for $\al=2.5$, and (d) $X$ point with
$(k_x,k_y)=(\pi/3 ,\pi/3\sqrt{3})$ for $\al=2$. Note that the
number of avoided crossings of the instantaneous eigenvalues can be
clearly read off from these figures; this number varies with $\vec
k$ as can be seen by comparing the plots from different panels. All
energies are in units of $\ga$.} \label{fig2} \end{figure}

A similar analysis charted out in the Appendix shows that for an odd
number ($2n+1)$ of level crossings, the phase bands
$\phi^{(2n+1)}(\vec k, t) \equiv \phi^o$ are given by
\begin{widetext}
\bea \cos (\phi^o) &=& \sum_{jmax=0}^{2n-1} \sum_{\al}
\prod_{j_{\al} = 1}^{jmax} (1-p_{j_{\al} \vec k})^{1/2}
\prod_{j'_{\al} \ne j_{\al}=1}^{2n-1-jmax} p_{j'_{\al} \vec
k}^{1/2} (-1)^{n_2} (1-\de_{n_{j'_{\al}},1})
\cos\left[\Phi_{\vec k}^s + \xi_{\vec k}^s (t) -\sum_a (\ga_a
\Phi_{a \vec k} + \de_a \xi_{a \vec k} (t)) \right], \label{oddcos1} \non \\
\eea
\end{widetext}
where $n_{j'_{\al}}$ denotes the number of occurrence of $j'$,
({\it i.e.}, the number of $\sqrt{p_{j_{\al}\vec k}}$ factors) in a
permutation $\al$, and $n_2= n_{j_{\al}}$ with $ {\rm
Max}[j'_{\al}]<j_{\al}<{\rm Min}[j'_{\al}]$.

Eqs.\ \eqref{evencos1} and \eqref{oddcos1} represent the main results of
this section. They allow us to chart out semi-analytic conditions
for phase band crossings and hence the topology change of a class of
driven integrable quantum models when the drive frequency is low
compared to the natural energy scale of the system Hamiltonian.
These conditions can be summarized as follows. For any $\vec k$,
these driven models, at low frequency, will exhibit a topology
change at a time $t_f$ during a drive which has been preceded by $n$
avoided level crossings of its instantaneous eigenvalues if
\bea \eta_{\vec k} &=& 1 ~~~{\rm and} ~~~\cos[\phi^{(n)}(\vec k,t_f)]= \pm 1,
\label{finalcond} \eea
where the expressions for $\cos[\phi^{(n)}(\vec k,t_f)]$ are given in
Eq.\ \eqref{evencos1} for even $n$ and Eq.\ \eqref{oddcos1} for odd $n$.
In the next section, we shall compare the predictions of these
equations for specific values of $n$ in the context of graphene.

\section {Phase diagram for graphene with external radiation}
\label{secphd}

In this section, we study the phase diagram of graphene in the
presence of external radiation in the low-frequency regime. In doing
so, we shall compare and contrast between the results obtained in
Sec.\ \ref{secadimp} and their counterparts from exact numerical
solution of the time-dependent Schr\"odinger equation which reads
(Eq.\ \eqref{graham1})
\bea i \hbar \frac{du_{\vec k} (t)}{dt} = Z_{\vec k} (t) v_{\vec k} (t), \quad
i \hbar \frac{dv_{\vec k} (t)}{dt} = Z^{\ast}_{\vec k} (t) u_{\vec k} (t).
\label{schro1} \eea
Such a comparative analysis between the adiabatic-impulse method and exact
numerics will serve to check the accuracy of the former in the
low-frequency regime. In what follows, we shall compare the two
methods in Sec.\ \ref{secphdcomp}. This will be followed by an
analysis of the phase band crossing conditions for several
high-symmetry points in the Brillouin zone of graphene in Sec.\
\ref{secphdsym}. Finally, based on the results obtained in these two
sections, we shall present the phase diagram of graphene under
external radiation in Sec.\ \ref{secphdpd}.

\subsection{Comparison between adiabatic-impulse and exact numerical
results} \label{secphdcomp}

In order to compare the results obtained by the adiabatic-impulse and exact
numerics, we focus on four representative points in the graphene Brillouin
zone and choose four representative amplitudes of radiation. These are (a)
$\Ga$ point with $(k_x, k_y)=(0,0)$ and $\al=2.35$, (b) Dirac ($K$) point
$(k_x,k_y)=(0,4 \pi/3 \sqrt{3})$ and $\al=2$, (c) $M$ point $(k_x,k_y)=
(2 \pi/3 , 0)$ with $\al=2.5$, and (d) $X$ point with $(k_x,k_y)=(\pi/3,
\pi/3\sqrt{3})$ with $\al=2$; these points are shown in Fig.\ \ref{fig1}.

\begin{figure}[t!]
\begin{center}
\includegraphics[width=\columnwidth]{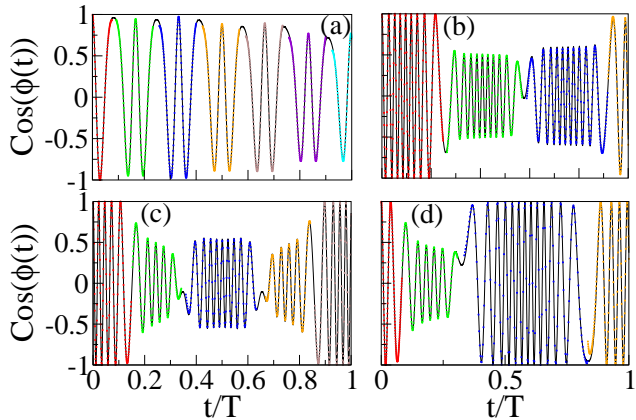}
\end{center}
\caption{Comparison of the phase bands $\cos(\phi_{\vec k} (t))$
obtained from exact numerics (solid lines) and adiabatic-impulse
approximation (dots) as a function of $t/T$ for (a) $\Ga$ point with
$\al=2.35$, (b) Dirac ($K$) point with $\al=2$, (c) $M$ point with
$\al=2.5$, and (d) $X$ point with $\al=2$. The different colors in
each plot correspond to different adiabatic regions separated by
avoided level crossings; see text for details.} \label{fig3}
\end{figure}

The expressions for the phase bands obtained from the adiabatic-impulse
approximation involve finding the number of avoided crossings
of the instantaneous energy eigenvalues within a given period. Thus
we first chart these out in Fig.\ \ref{fig2} for the above-mentioned
points in the Brillouin zone. In addition, we also show the
wave function overlap $\eta_{\vec k}(t)$ for each of these points.
The symmetry of the instantaneous eigenvalues becomes clear from
these plots; these will be discussed in Sec. \ref{secphdsym}. Here
we note that after obtaining the number of such avoided crossings
for the specified $\vec k$ values, we can use the results of Sec.\
\ref{secadimp} to obtain the evolution matrix $U_{\vec k}(t,0)$. We
then diagonalize $U_{\vec k}(t,0)$ to obtain the expressions for the
phase bands as given by the adiabatic-impulse method. This can be
done either analytically for a small number of crossings ($n <3$)
using Eqs.\ \eqref{pbreg1}, \eqref{pbreg2} and \eqref{pbreg3} or numerically
using Eqs.\ \eqref{etadef}, \eqref{uadrel1}, and \eqref{phasebands1}.

These results for the phase bands obtained using the
adiabatic-impulse method is then compared against their numerical
counterparts. The latter procedure involves a numerical solution of the
Schr\"odinger equation (Eq.\ \eqref{schro1}), followed by a construction of
$U_{\vec k} (t)$ from the final wave function using (Ref. \onlinecite{asen1}),
\bea \left(\begin{array}{c} u_{\vec k} (t) \\ v_{\vec k} (t) \end{array}
\right) &=& U_{\vec k}(t,0) \left(\begin{array}{c} u_{\vec k} (0) \\
v_{\vec k} (0) \end{array} \right). \label{uconst1} \eea
Finally, we numerically diagonalize the matrix $U_{\vec k}(t,0)$ for various
values of $\vec k$. This leads to numerically exact expressions for the
phase bands.

\begin{figure}[t!]
\begin{center}
\includegraphics[width=\columnwidth]{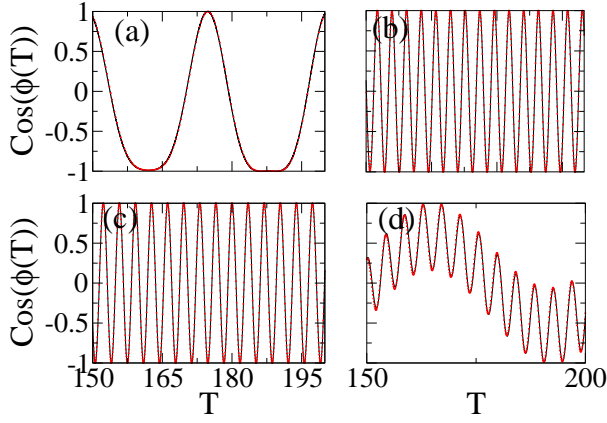}
\end{center}
\caption{Comparison of the phase bands
$\cos(\phi_{\vec k} (T))$ obtained from exact numerics (black solid
lines) and adiabatic-impulse approximation (dotted lines) as a
function of $T$ (in units of $\hbar/\ga$). All parameters are the
same as in Fig.\ \ref{fig2}.}\label{fig4} \end{figure}

A comparison between $\cos(\phi_{\vec k} (t))$ obtained using these
two methods, shown in Fig.\ \ref{fig3}, shows a near-exact match;
this indicates that the adiabatic-impulse method reproduces the
phase bands accurately for all $t \le T$. We have numerically
checked this result for several other amplitudes and Brillouin zone
points; the phase bands obtained via the adiabatic-impulse
approximation always shows a very good agreement with its exact
numerical counterpart. A similar comparative plot of
$\cos(\phi_{\vec k} (T))$, shown in Fig.\ \ref{fig4}, indicates a
near-exact match of these bands for a wide range of drive
frequencies. We have numerically verified that the expressions for
the phase band obtained using the adiabatic-impulse method matches its
numerical counterparts for all $\om \le 1$ and for $\al \le
5$. Thus the adiabatic-impulse approximation seems to be accurate at
low drive frequencies; we shall use this fact in Sec.\
\ref{secphdsym} to analyze the phase band crossings of graphene.

\subsection{Phase band crossing conditions}
\label{secphdsym}

In this section, we provide explicit phase band crossing
conditions for several high-symmetry points in the graphene
Brillouin zone and check their validity by comparison with exact
numerical results. This will be followed by general comments about
generic $\vec k$ points in the Brillouin zone.

{\it $\Ga$ point}: As shown in Sec.\ \ref{secsymm}, the graphene
Hamiltonian exhibits $T/3$ periodicity at the $\Ga$ point since
$H_{\vec k} (t)= H_{\vec k}(t + T/3)$ for $\vec k = (0,0)$ (Eq.\
\eqref{symgam3}). Thus we only need to track the evolution for
$t \le T/3$ for identifying the phase band crossings. Furthermore, as shown
in Fig.\ \ref{fig2} (a) and derived in Sec.\ \ref{secsymm} (Eq.\
\eqref{symgam2}), the instantaneous energy eigenvalues $E_{\vec k=(0,0)} (t)
\equiv E_\Ga (t)$ satisfy $E_\Ga (nT/6 \pm t)= E_\Ga (t)$ for any
integer $n$. Thus the kinematic phase, $\chi_\Ga (t,0)$, picked up by the
system at the $\Ga$ point during the adiabatic evolution satisfies
\bea \xi_\Ga (t,0) &=& \xi_\Ga (t\pm nT/6,0), \non \\
\xi_{1 \Ga} &=& \xi_{3 \Ga}= \xi_{2 \Ga}/2 =\xi_\Ga,
\label{gammacond1} \eea where in the last line we have used the
notation $\xi_\Ga (t_{i \Ga}, t_{i-1 \Ga}) \equiv \xi_{i
\Ga}$, and $t_{i \Ga}$ denotes the time of the $i^{\rm th}$
avoided level crossing at the $\Ga$ point. From Fig.\ \ref{fig2}
(a), we find that $t_{1\Ga}=T/12$ and $t_{2\Ga}=T/4 =
t_{1\Ga}+T/6$ which leads to Eq.\ \eqref{gammacond1}. Further, it
is easy to check that $|dE_{\Ga}(t)/dt|$ satisfies
$|dE_{\Ga}(t+nT/6)/dt|=|dE_\Ga (t)/dt|$; using this it is
possible to check that the Landau-Zener excitation probabilities
$p_{i \vec k=(0,0)} \equiv p_{i \Ga}$ for $i=1,2$ (corresponding
to times $t_{1\Ga}=T/12$ and $t_{2\Ga}=T/4$) and the
corresponding Stuckelberg phases $\Phi_{i \vec k=(0,0)} \equiv
\Phi_{i \Ga}$ at the $\Ga$ point satisfy (Eqs.\ \eqref{lzprob}
and \eqref{crel1}) \bea p_{1\Ga} &=& p_{2\Ga} =p_\Ga ,
\quad \Phi_{1\Ga}= \Phi_{2\Ga}= \Phi_\Ga.
\label{gammacond2} \eea Using Eqs.\ \eqref{gammacond1} and
\eqref{gammacond2}, we then obtain an expression for the phase bands
(Eq.\ \eqref{pbreg3sim}) at $t=T/3$ where $\eta =1$ as
\bea \cos(\phi_\Ga (T/3)) &=& p_\Ga + (1-p_\Ga ) \cos(2\Lambda_\Ga ),
\label{pbandgamma1} \eea
where $\Lambda_\Ga = \Phi_\Ga + 2\xi_\Ga $.

\begin{figure}[t!]
\begin{center}
\includegraphics[width=\columnwidth]{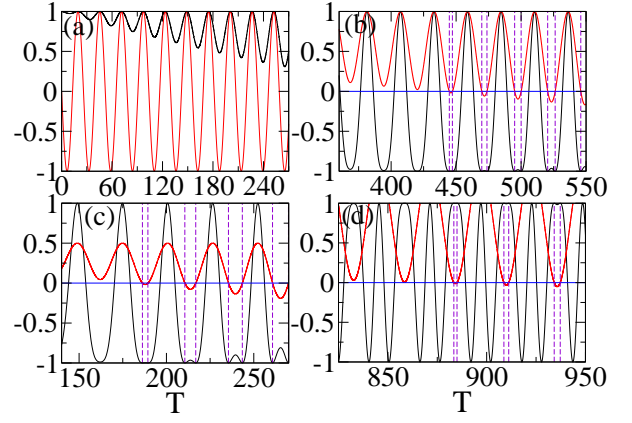}
\end{center}
\caption{Comparison of phase band crossing conditions
from adiabatic-impulse approximation (Eqs.\ \eqref{gammacrosscond1},
\eqref{gammacrosscond2}, and \eqref{gammacrosscond3} with exact
numerics. In all plots $t$ and $T$ are in units of $\hbar/\ga$.
(a) Phase band crossings at $t=T/3$ from Floquet zone center
($\phi_\Ga (T/3)=2 n \pi$). Black solid line indicates numerical
plot of the phase bands at $t=T/3$ as a function of $T$. The red
dotted lines show a plot of $\cos[2(\Phi_\Ga +\xi_\Ga )]$
whose value is predicted to be unity when Eq.\
\eqref{gammacrosscond1} is satisfied. (b) Plot of phase band
crossings which occur at $t=2T/3$ through the Floquet zone edge
($\phi_\Ga (2T/3)= (2n+1)\pi$). The black solid line indicates
$\cos(\phi_\Ga (2T/3)$ while the red dotted lines show the left
side of Eq.\ \eqref{gammacrosscond2} which touches zero at all phase
band crossings in accordance with the adiabatic-impulse prediction.
(c) Phase band crossings at $t=T$ through the Floquet zone edge for
which $\phi_\Ga (T/3)= \pi/3$. The black solid line indicates
$\cos(\phi_\Ga (2T/3)$ while the red dotted lines show the left
side of Eq.\ \eqref{gammacrosscond3} which touches zero at each
phase band crossing. (d) Same as (c) but for $\phi_\Ga (T/3)=
2\pi/3$ for which the crossing condition is given by Eq.\
\eqref{gammacrosscond4}.} \label{fig5} \end{figure}

Eq.\ \eqref{pbandgamma1} can be used to relate the phase band
crossing condition to the Landau-Zener probability $p_\Ga $ and the
Stuckelberg phase $\Phi_\Ga $. To see this, we first note that a
class of such crossings occurs at the Floquet zone center (namely,
$\cos(\phi_\Ga (T/3))=1$, so that $\phi_\Ga (T/3) = 2 \pi n$ at the
crossing). From Eq.\ \eqref{pbandgamma1}, we find that the condition
for such crossings is given by \bea \Lambda_\Ga = m \pi, ~~~~{\rm
where}~~~~ m \in Z. \label{gammacrosscond1} \eea We would like to
point out here that the condition $\phi_\Ga (T/3)= (2 n+1) \pi$ is
untenable since this would require $ \cos[2(\Phi_\Ga + \xi_\Ga )] =
-(1+p_\Ga )/(1-p_\Ga ) < -1$ for any $p_\Ga >0$. Thus our analysis
predicts that all phase band crossings at the $\Ga$ point which
occur at $t=T/3$ must be through the Floquet zone center. Note that
all such crossings through zone center at $t=T/3$ also imply
corresponding crossings at $t=2T/3$ and $t=T$ as can be seen from
Eq.~\eqref{ugammacond1}.

The second class of crossings that occurs at the $\Ga$ point
involves a phase band crossing at $t=T$ or $t=2T/3$ without an
analogous crossing at $t=T/3$. Such crossings, at $t=2T/3$, always
occur through the Floquet zone edges ($\cos(\phi_\Ga) =-1$, so that
$\phi_\Ga = (2n+1)\pi$ at the crossing), and they can be understood
as follows. First, we note that from Eq.\ \eqref{ugammacond1}, we
have $\phi_\Ga (T/3) = \phi_\Ga (2T/3)/2= \phi_\Ga (T)/3$. Thus it
is possible to have a phase band crossing at $t=2T/3$ without any
crossing at $t=T/3$ if $\phi_\Ga (T/3)= \pi/2$. This requires \bea
\cos(2\Lambda_\Ga )+ p_\Ga /(1-p_\Ga ) &=& 0.
\label{gammacrosscond2} \eea Note that such crossings can occur for
$p_{\Ga} \le 1/2$; further for any $p_{\Ga} < 1/2$ we expect two
solutions to Eq.\ \eqref{gammacrosscond2} leading to a pair of
possible crossings for a given $p_{\Ga}$. Second, for $t=T$, we can
have similar crossings for which $\phi_\Ga (T/3)= \pi/3$ and
$\phi_\Ga (T/3)=2\pi/3$ which lead to the conditions \bea
\cos(2\Lambda_\Ga ) - \frac{1-2p_\Ga }{2(1-p_\Ga )} &=& 0 , \,\,
{\rm for}\,\, \phi_\Ga (T/3)= \frac{\pi}{3}, \label{gammacrosscond3} \\
\cos(2\Lambda_\Ga ) + \frac{1+2p_\Ga }{2(1-p_\Ga )} &=& 0, \,\, {\rm
for}\,\, \phi_\Ga (T/3)= \frac{2\pi}{3}. \label{gammacrosscond4}
\eea We note Eq.\ \eqref{gammacrosscond3} predicts that such
crossings through Floquet zone edge can only occur for $p_{\Ga} \le
3/4$, and for $p_{\Ga} < 3/4$, there are a pair of crossings for a
given value of $p_{\Ga}$. In contrast, Eq.\ \eqref{gammacrosscond4}
shows that for $\phi_{\Ga}=2 \pi/3$, crossings occur through the
Floquet zone center in pairs if $p_{\Ga} \le 1/4$. The crossing
conditions charted out in Eqs.\ \eqref{gammacrosscond2} and
\eqref{gammacrosscond3} constitute an example of the response of the
system at fractional frequencies $2\om/3$ ( Eq.\
\eqref{gammacrosscond2}) and $\om/3$ (Eqs.\
\eqref{gammacrosscond1}); these occur since the evolution operator
$U$ does not have the same periodicity as $H$ at these drive
frequencies.

A comparison between the crossing conditions obtained above and the
exact numerical result is shown in Fig.\ \ref{fig4}. The top
panel of this figure shows that all the phase band crossings for
$t=T/3$ at the $\Ga$ point is consistent with Eq.\
\eqref{gammacrosscond1} for a wide range of $T \ge 2 \pi$ as shown in
Fig.\ \ref{fig5}. We find that the analytic condition presented in
Eqs.\ (\ref{gammacrosscond1}-\ref{gammacrosscond4}) is exactly
satisfied for all the phase band crossings that we find using exact
numerics. Furthermore, we do not find any phase band crossings at
$t=T/3$ which occur through the Floquet zone edge which is
consistent with our expectation from the adiabatic-impulse theory.

{\it Dirac points}: For the Dirac points, we find from Fig.\
\ref{fig2} (b) that there are three avoided level crossings
($j=1,2,3$) which divide the evolution into four regions denoted as
$i=1, 2, 3, 4$. We have confirmed numerically that the number of such
crossings does not change for $\al \le 5$ and our subsequent
discussions will hold in this regime. Also, in what follows, we shall
explicitly study the $K$ point for which $(k_x, k_y)=(0, 4 \pi/(3
\sqrt{3}))$; all our results will also hold for the $K'$ point.

\begin{figure}[t!]
\begin{center}
\includegraphics[width=\columnwidth]{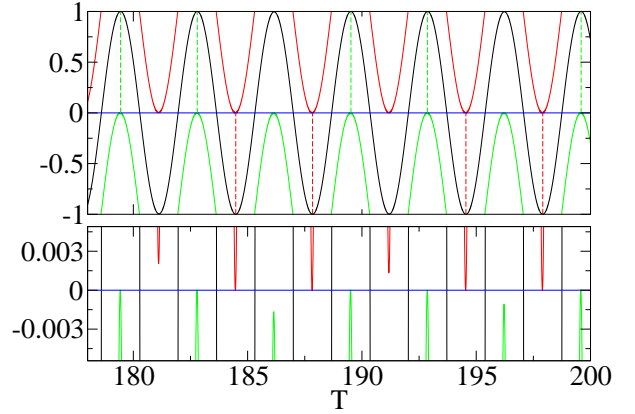}
\end{center}
\caption{Comparison of phase band crossing conditions from
adiabatic-impulse approximation (Eq.\ \eqref{diraccrosscond}) with
exact numerics at the Dirac point $K$. Top panel: The black solid
line shows the numerical phase band ($\cos(\phi_K)$) as a function
of $T$. The red (green) lines plots the left side of Eq.\
\eqref{diraccrosscond}; these touch zero (and hence satisfy Eq.\
\eqref{diraccrosscond}) for crossings through the Floquet zone edge
(center). The vertical lines are guides to the eye for the positions
of exact crossings; these lines are absent when the crossings are
avoided. Bottom panel: A zoomed plot of the left side of Eq.\
\eqref{diraccrosscond} showing that these curves do not touch zero
at the positions of avoided phase band crossings. In both plots $T$
is in units of $\hbar/\ga$.} \label{fig6} \end{figure}

The symmetries of the instantaneous Hamiltonian $H_K (t)$ at the
$K$ point are listed in Eqs.\ (\ref{symdir1}-\ref{symdir5}).
Using these conditions, it is easy to see that $E_K (t)= E_K(t+T/3)$
for all $t$. Thus the kinematic phase $\xi_{i K} \equiv \xi_{(k_x,
k_y)=(0, 4\pi/(3\sqrt{3}))} (t_{iK}, t_{i-1 K})$ picked up in the
$i^{\rm th}$ region satisfies
\bea \xi_{2K} &=& \xi_{3K}= \xi_K,\,\, \xi_{1K}= \xi'_K, \,\, \xi_{4K}=
\xi_K-\xi'_K. \label{kinphaseK} \eea
Furthermore since the avoided level crossings occur at time
differences $\Delta t= T/3$, we find that the Landau-Zener
probabilities $p_{j K}$ for $j=1, 2, 3$ and the corresponding
Stuckelberg phases $\Phi_{j K}$ at each of the avoided crossings are
identical. We therefore denote
\bea p_{j K} &=& p_K ~~~{\rm and}~~~ \Phi_{j K}= \Phi_K. \label{stuckK} \eea
Using these symmetries and following the method charted out in the
Appendix, we find that the expression for the phase band at $t=T$
where $\eta_K=1$ is given by
\bea \cos \phi_K (T) &=& \sqrt{1-p_K} [ (1-p_K) ~\cos (3\Lambda_K) \non \\
&& ~~~~~~~~~~~~~~~~~~-~ 3 p_K ~\cos\Lambda_K ], \label{pbandK} \eea
where $\Lambda_K= \xi_K + \Phi_K$. The conditions for phase
band crossings at the Dirac point, through the Floquet zone center
($\cos[\phi_K (T)]=1$) or the Floquet zone edge ($\cos[\phi_K (T)]=-
1$), are therefore given by
\bea \mp \frac{1}{\sqrt{1-p_K}} + \cos (3\Lambda_K) - 4 p_K \cos^3 \Lambda_K
&=& 0, \label{diraccrosscond} \eea
where the upper (lower) sign corresponds to crossings through
the Floquet zone center (edge). We note that Eq.\
\eqref{diraccrosscond} predicts that no phase band crossings occur for
$p_K > 3/4$. A plot of the left side of Eq.\ \eqref{diraccrosscond}
as a function of $T$ is shown in Fig.\ \ref{fig6}. We find that
these curves touch zero and hence satisfy Eq.\
\eqref{diraccrosscond} precisely at the locations of the phase band
crossings as predicted by exact numerics.

\begin{figure}[t!]
\begin{center}
\includegraphics[width=\columnwidth]{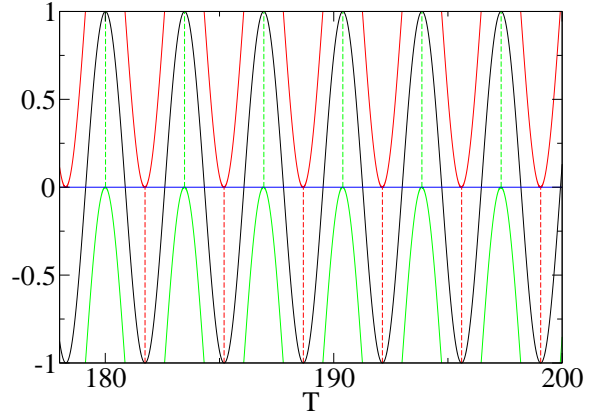}
\end{center}
\caption{Comparison of phase band crossing conditions from
adiabatic-impulse approximation (Eq.\ \eqref{Mcrosscond}) with exact
numerics at the $M$ point. Top panel: The black solid line shows the
numerical phase band ($\cos(\phi_M)$) as a function of $T$. The red
(green) lines are plots of the left side of Eq.\ \eqref{Mcrosscond};
these touch zero [and hence satisfy Eq.\ \eqref{Mcrosscond}] for
crossings through the Floquet zone edge (center). The vertical lines
are guides to the eye for the positions of exact crossings. $T$ is
in units of $\hbar/\ga$.} \label{fig7} \end{figure}

{\it $M$ point}: At the $M$ point for which $(k_x, k_y)=
(2\pi/3,0)$, we find, from Fig.\ \ref{fig2} (c), four avoided level
crossings corresponding to $j=1, 2, 3, 4$, which divide the
evolution into five adiabatic regions denoted by $i=1, 2, \cdots,
5$. The symmetries of the instantaneous Hamiltonian have been listed
in Eqs.\ \eqref{symcor1} and \eqref{symcor2}. These symmetries
ensure that at the $M$ point $E_M(T-t)=E_M (t)$ and
$E_M(T/2+t)=E_M(T/2-t)=E (t)$. The kinematic phases picked up in the
adiabatic regions $\xi_{i M} \equiv \xi_M(t_i, t_{i-1})$ for $i= 1,
2, \cdots, 5$ thus satisfy \bea \xi_{2M}= \xi_{4M}= \xi'_M, \quad
\xi_{1M}= \xi_{5M}= \xi_{3M}/2 =\xi_M. \label{kinphaseM} \eea
Further, since the four avoided crossings occur at times $t_0$,
$T/2-t_0$, $t_0 + T/2$, and $T-t_0$ where $t_0 \simeq 0.15 T$, they
have the same Landau-Zener probabilities and Stuckelberg phases: we
therefore denote $p_{jM}= p_M$ and $\Phi_{jM}= \Phi_M$. Using these
symmetries, we can compute the phase bands, following the method
outlined in Sec.\ \ref{secadimp} and the Appendix, to obtain \bea &&
\cos \phi_M (T) ~=~ (1- p_M)^2 \cos (2\Lambda_{+M}) +p_M^2 \cos
(2 \Lambda_{-M}) \non \\
&& -2p_M(1-p_M) [\cos (2\Lambda_{1M}) + \cos (2\Lambda_{2M}) + 1],
\label{pbandM} \eea where $\Lambda_{1M}= \Phi_M + 2 \xi_M$,
$\Lambda_{2M}= \Phi_M+\xi'_M$ and $\Lambda_{\pm M}= \Lambda_{1M} \pm
\Lambda_{2M}$. This leads to the phase band crossing conditions \bea
&& \mp \frac{1}{(1-p_M)^2} +\cos(2\Lambda_{+M}) +\frac{p_M^2}{
(1-p_M)^2} \cos (2\Lambda_{-M}) \non \\ && - \frac{2p_M}{1-p_M}
[\cos (2\Lambda_{1M}) + \cos (2\Lambda_{2M}) + 1] =0,
\label{Mcrosscond} \eea where the upper (lower) sign holds for phase
band crossing through the Floquet zone center (edge) [$\cos(\phi_M
(T))=1 (-1)$]. Numerically, we find that Eq.\ \eqref{Mcrosscond} can
be satisfied for all $p_M \le 0.9492116$. A plot of the left side of
Eq.\ \eqref{Mcrosscond} is shown in Fig.\ \ref{fig7}; these curves
are found to touch zero (and hence satisfy Eq.\ \eqref{Mcrosscond})
at the positions of the phase band crossings obtained from exact
numerics.

\begin{figure}[t!]
\begin{center}
\includegraphics[width=\columnwidth]{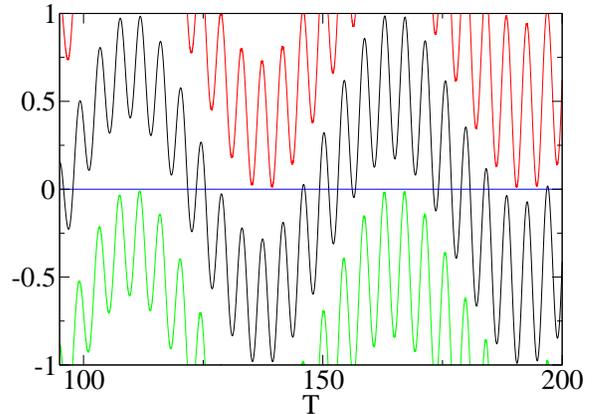}
\end{center}
\caption{Comparison of phase band crossing conditions
from adiabatic-impulse approximation (Eq.\ \eqref{xcrosscond}) with
exact numerics at the $X$ point. The black solid line shows the
numerical phase band ($\cos(\phi_M)$) as a function of $T$. The red
(green) lines are plots of the left side of Eq.\ \eqref{xcrosscond}.
We do not find any phase band crossings for $\al=2$ and as predicted, these
lines never touch zero indicating that Eq.\ \eqref{Mcrosscond} is never
satisfied. $T$ is in units of $\hbar/\ga$.} \label{fig8} \end{figure}

{\it $X$ point}: As shown in Fig.\ \ref{fig2} (d), for the $X$ point where
$(k_x, k_y)= (\pi/3, \pi/((3 \sqrt{3}))$, we have, for $\al \le 2.5$, three
avoided level crossings corresponding to $j=1, 2, 3$, which divide the
evolution into four adiabatic regions which we label as $i=1, 2, 3, 4$.
In this case, using Eq.\ \eqref{symx1}, we find $E_{\vec k} (t)=
E_{\vec k}(T/6-t)= E_{\vec k}(7T/6-t)$. Using this we can establish that the
kinematic phases in regions $1, 2, 3$ and $4$ satisfy
\bea \xi_{2X} &=& \xi_{1X}+\xi_{3X}= \xi_X, ~~~{\rm and}~~~ \xi_{3X}= \xi'_X.
\label{kinphasex} \eea
Further, since the avoided crossings corresponding to $j=2$ and $j=3$
occur at $t= t_0 \simeq 0.3 T$ and $t \simeq 0.85 T = 7T/6-t_0$, we have
\bea p_{2X} &=& p_{3X}= p_X, ~~~{\rm and}~~~ p_{1X}= p'_X, \non \\
\Phi_{2X} &=& \Phi_{3X}= \Phi_X, ~~~{\rm and}~~~ \Phi_{1X}= \Phi'_X.
\label{probx} \eea
Using these symmetries and following the method charted out in the
Appendix we obtain, at $t=T$ where $\eta_X=1$,
\bea && \cos \phi_X (T) = \sqrt{1-p'_X}(1-p_X) \cos(2\Lambda_{1X}+\Lambda_{2X})
\non \\
&& -2 \sqrt{p_X p'_X(1-p_X)} \cos\Lambda_{1X} - p_X \sqrt{1-p'_X}
\cos \Lambda_{2X}, \non \\
&& \label{bandX} \eea
where $\Lambda_{1X}= \Phi_X+\xi'_X$ and $\Lambda_{2X}=\Phi'_X+2\xi_X-\xi'_X$.
Thus the phase band crossing condition is given by
\bea &&\pm ~\frac{1}{\sqrt{1-p'_X}(1-p_X)} ~-~ \cos(2\Lambda_{1X}+\Lambda_{2X})
\non \\
&& + ~2 ~\sqrt{\frac{p_X p'_X}{(1-p_X)(1-p'_X)}} ~\cos\Lambda_{1X} \non \\
&& + ~\frac{p_X}{1-p_X} \cos \Lambda_{2X} ~=~ 0, \label{xcrosscond} \eea
where the $+ (-)$ sign corresponds to a phase band crossing
through the zone center (edge). A plot of the left side of Eq.\
\eqref{xcrosscond} is shown in Fig.\ \ref{fig8} for $\al=2$; we
find that they never touch zero within the range of $T$ shown in the
figure. This is consistent with the fact that there are no phase
band crossings within this range as can be seen from a plot of
$\cos(\phi_X (T))$; in fact, we have numerically checked that for
$\al=2$, the phase bands do not cross for any $T \ge 2 \pi$.
Numerically, we find very few such crossings which will be discussed
in Sec.\ \ref{secphdpd}.

The decrease in the number of phase band crossings at the $X$ point
is a consequence of the lack of a symmetry necessary to make all
$p_{iX}$ and $\Phi_{iX}$ equal; in fact such a reduction resonates
with the fact that phase band crossings are much more difficult to
find at arbitrary points in the Brillouin zone which are not equal
to any of the high symmetry points. The crucial role of symmetry for
phase band crossings can be understood as follows. Consider a case
of two avoided level crossings each having a probability $p_i$ and a
Stuckelberg phase $\Phi_i$ for $i=1,2$. Let the phases picked up in
the corresponding adiabatic regions I, II and III to be $\xi_j$
where $j=1,2,3$. Then for a situation with no symmetry, the phase
bands are given by (Eq.\ \eqref{pbreg3})
\bea \cos (\phi) &=& \sqrt{(1-p_1)(1-p_2)} \cos(\sum_{i=1,2} \phi_i +
\sum_{i=1,3} \xi_i) \non \\
&& +\sqrt{p_1 p_2} \cos(\xi_1+\xi_3-\xi_2) \label{nosym1} \eea Now
we look for possibility of tuning $\cos (\phi)= \pm 1$ by varying
$\al$ and $T$. Since $\Phi_i$ is essentially a function of $p_i$, we
have five independent quantities, $p_{1,2}$ and $\xi_{1,2,3}$ to
tune by varying $\al$ and $T$. This requires fine tuning. In
contrast, in the presence of symmetries as in the case of the $\Ga$
point, where $p_1=p_2 = p_{\Ga}$ and $\xi_1=\xi_2/2=\xi_3
=\xi_{\Ga}$, one has to just tune two parameters by varying $\al$
and $\Ga$. This does not require fine tuning. The argument easily
extends to a larger number of crossings since the number of
quantities to tune increases rapidly with the number of crossings in
the absence of any symmetry. Thus, we generally expect phase band
crossings to happen only at $\vk$ points which have the requisite
symmetries to make most $p_i$ and $\xi_i$s equal. We note that our
numerical search in the graphene Brillouin zone confirms this
expectation. We shall not discuss the non-generic phase band
crossings at arbitrary low-symmetry Brillouin zone points further in
this work.

\subsection{Phase diagram}
\label{secphdpd}

\begin{figure}[t!]
\begin{center}
\includegraphics[width=\columnwidth]{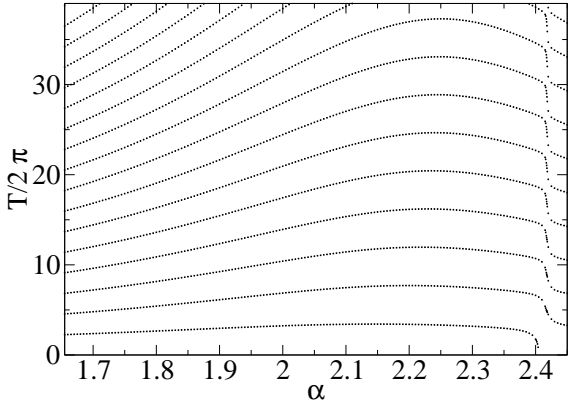}
\end{center}
\caption{Phase diagram at $t=T/3$ as a function of radiation
amplitude $\al$ and time period $T$ (in units of $\hbar/\ga$). Note that there
is a critical time period $T(\al)$ below which no phase band crossings occur
showing that there is no change in the topology of the phase bands at $t=T/3$
at high frequencies, $\om/\ga \gtrsim 1$. } \label{fig9} \end{figure}

\begin{figure}[t!]
\begin{center}
\includegraphics[width=\columnwidth]{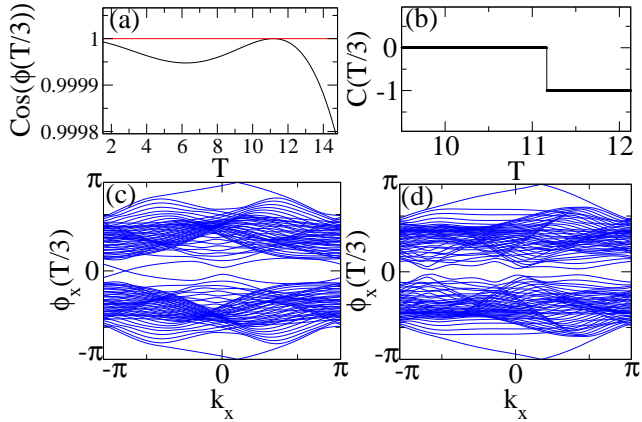}
\end{center}
\caption{Plot of change in Chern number and number of edge
states for $\al=2.4$ and $T=11.2$ (in units of $\hbar/\ga$) (a) Position of
the phase band crossing for which the Chern number is computed. (b) Change in
Chern number of the system across the phase band crossing at $T=T_c$ showing
$\Delta C=-1$. (c) and (d) The edge state structure around $E=0, \pi/T$ before
($T < T_c$) and after ($T >T_c$) the phase band crossing. The number of edge
states decreases by 1 across the transition as can be seen by comparing the
number of such states at $E=0$ in (c) and (d).} \label{fig10} \end{figure}

\begin{figure}[t!]
\begin{center}
\includegraphics[width=\columnwidth]{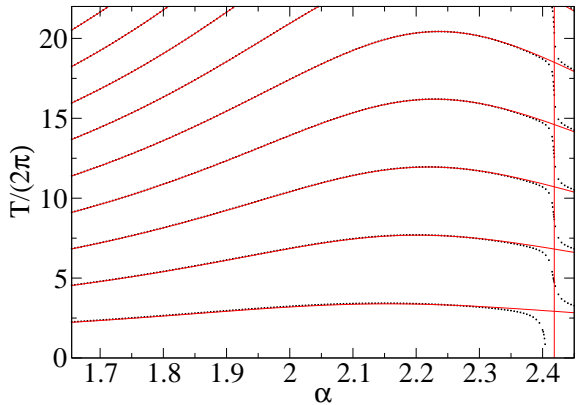}
\end{center}
\caption{Comparison of the exact numerical (black
dots) with adiabatic-impulse results (red dots) for phase band
crossings at $t=T/3$ (in units of $\hbar/\ga$) in the $\al-T$ plane. The red
solid line corresponds to the $p_{\Ga}=1$ line in the $\al-T$ plane. All
parameters are the same as in Fig.\ \ref{fig9}.} \label{fig11} \end{figure}

\begin{figure}[t!]
\begin{center}
\includegraphics[width=\columnwidth]{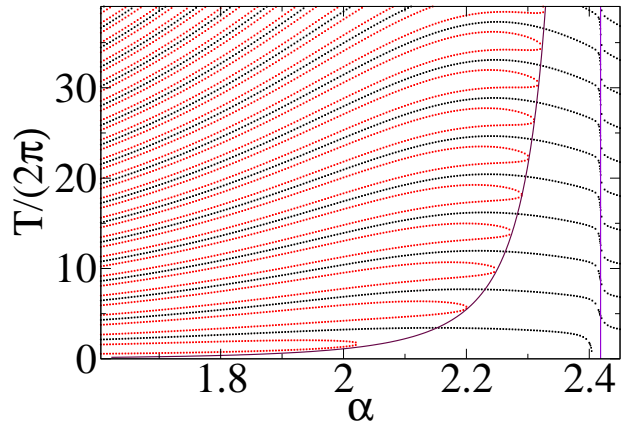}
\end{center}
\caption{The phase diagram as a function of $\al$ and $T$ (in units
of $\hbar/\ga$) at $t=2T/3$. The red dashed (black dotted) lines
indicate phase band crossing through the zone edge (center). The
solid lines indicate the curves $p_{\Ga}=1/2$ (brown solid line) and
$p_{\Ga}=1$ (violet solid line) in the $\al-T$ plane.} \label{fig12}
\end{figure}

\begin{figure}[t!]
\begin{center}
\includegraphics[width=\columnwidth]{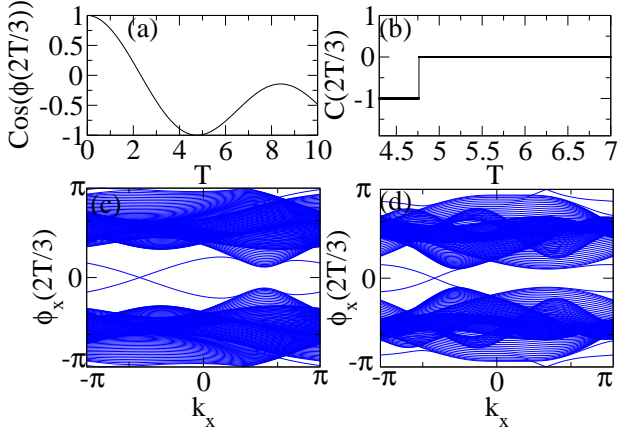}
\end{center}
\caption{Computation of the Chern number and number of edge states
for $t=2T/3$ at $\al=1.8$ and $T=4.76$ (in units of $\hbar/\ga$).
(a) Position of the phase band crossing for which the Chern number
is computed. (b) Change in Chern number of the system across the
phase band crossing at $T=T_c$ showing $\Delta C=1$. (c) and (d) The
edge state structure around $E=0$ and $\pi/T$ before ($T < T_c$) and
after ($T >T_c$) the phase band crossing. The number of edge states
increases by 1 across the transition as can be seen by comparing the
number of such states at $E=\pi/T$ in (c) and (d).} \label{fig13}
\end{figure}

\begin{figure}[t!]
\begin{center}
\subfigure[]{\includegraphics[width=0.45 \columnwidth]{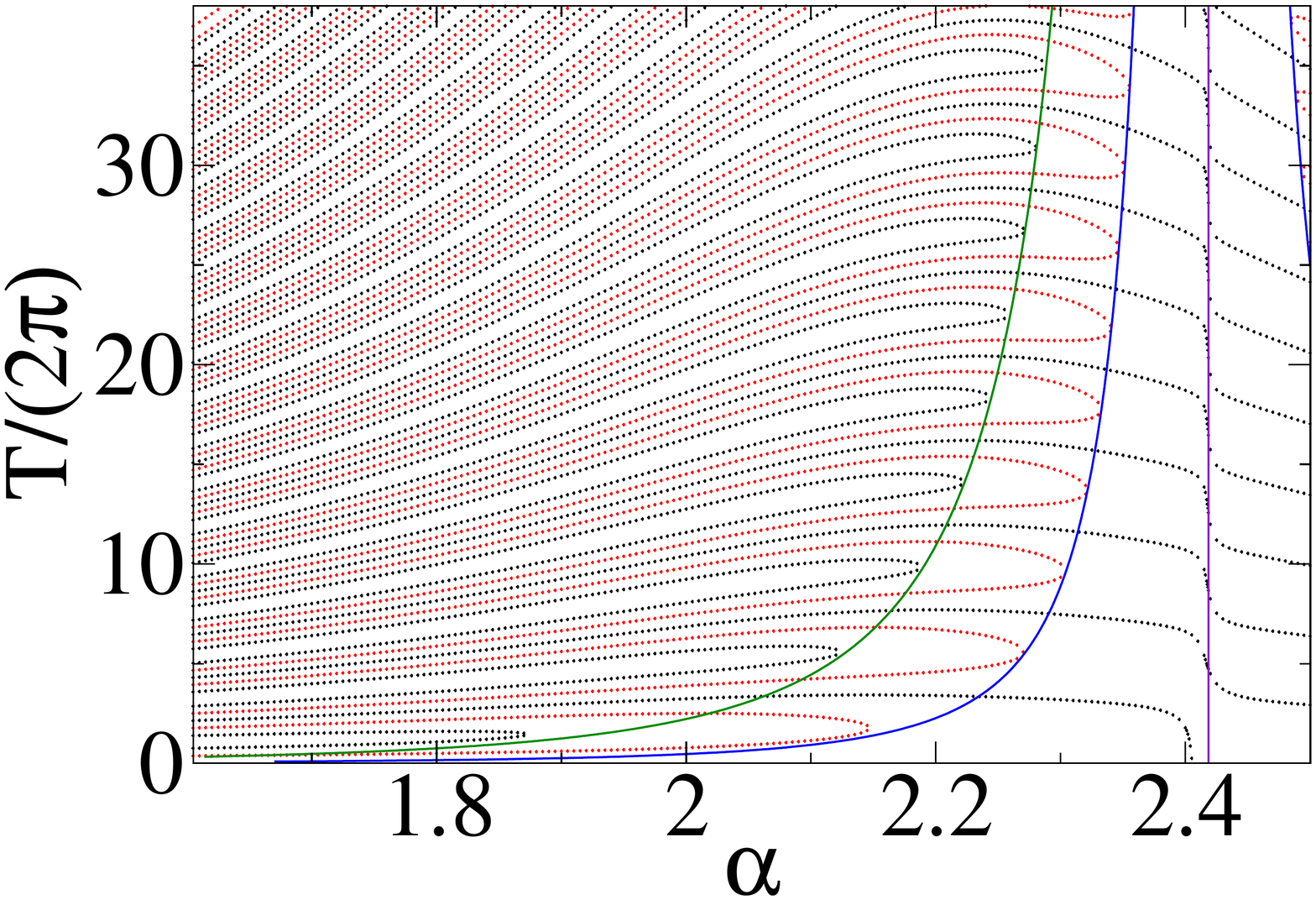}}
\subfigure[]{\includegraphics[width=0.45 \columnwidth]{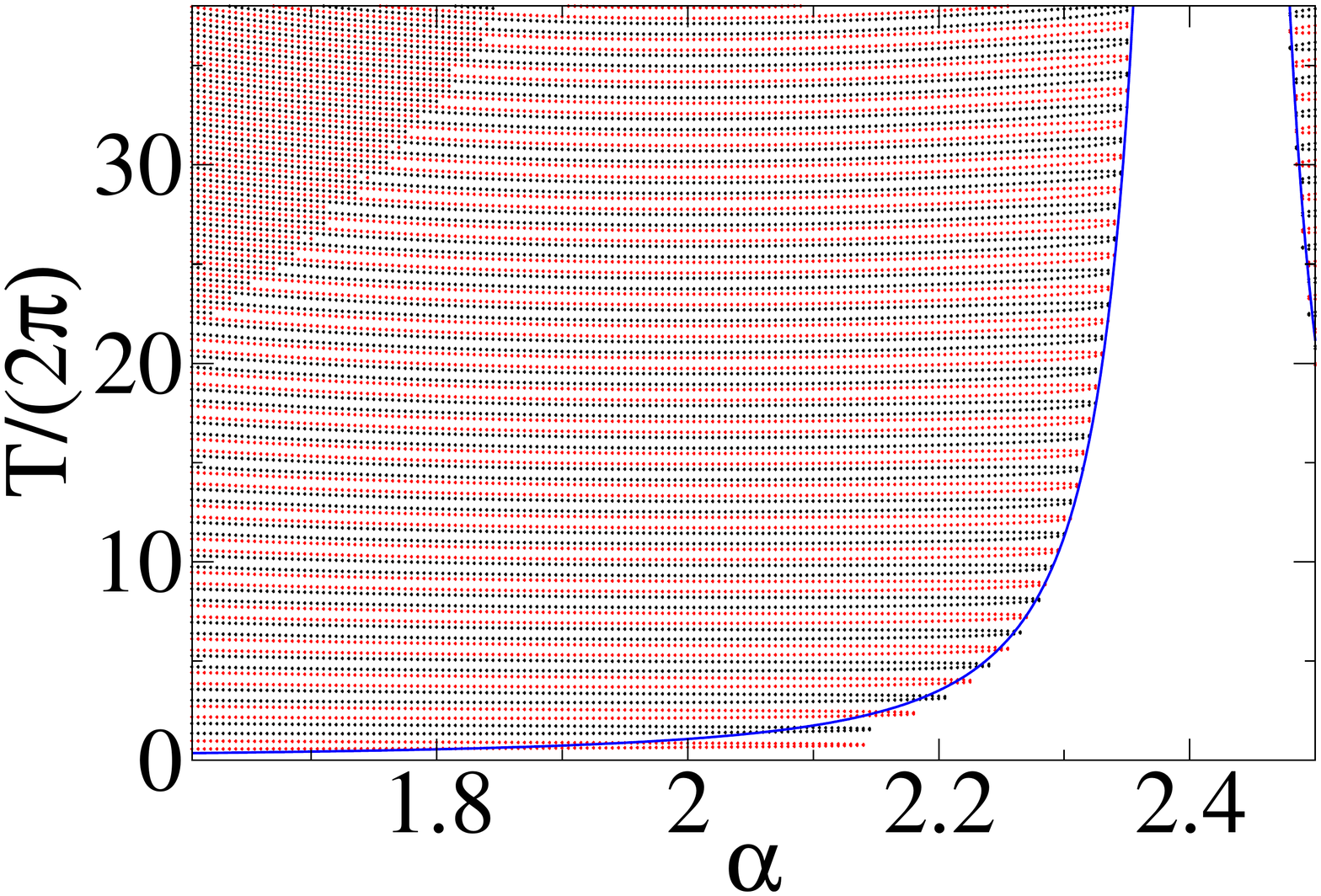}} \\
\subfigure[]{\includegraphics[width=0.45 \columnwidth]{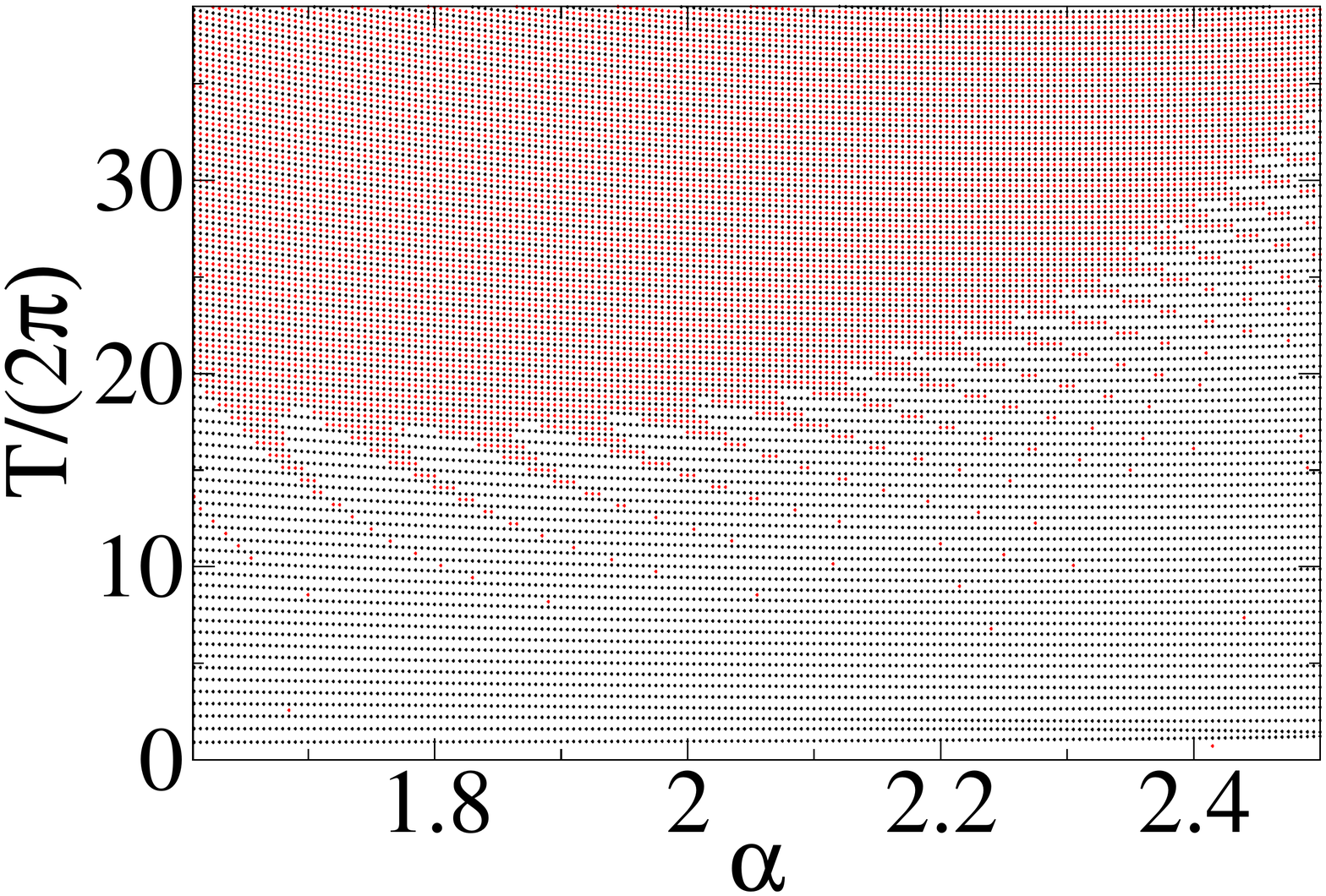}}
\subfigure[]{\includegraphics[width=0.45 \columnwidth]{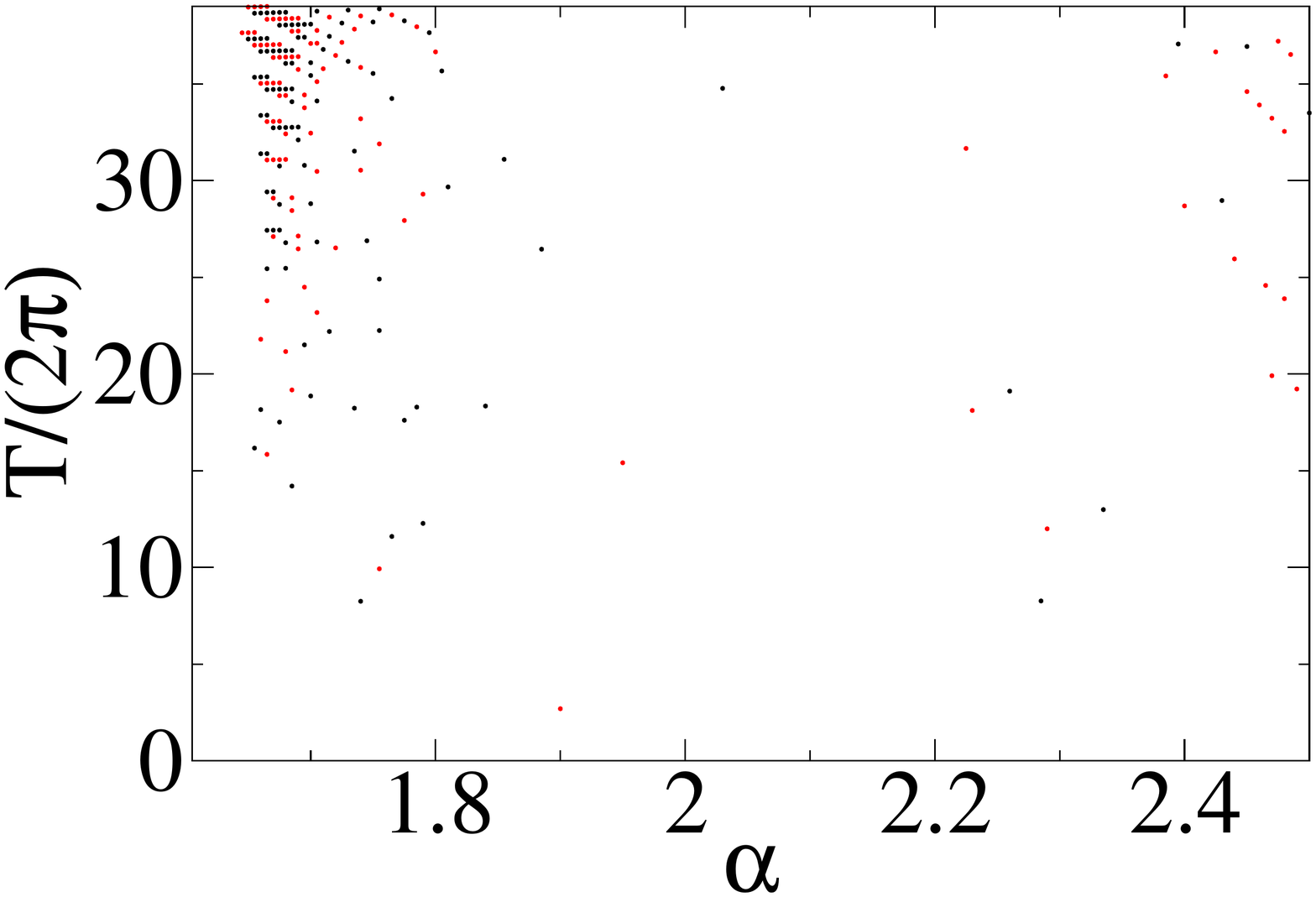}} \\
\end{center}
\caption{Contributions to the phase diagram at $t=T$ in the $\al-T$
plane from (a) $\Ga$, (b) $K$, (c) $M$ and (d) $X$ points. All
crossings through the zone center (edge) are indicated by black
(red) dots, and $T$ is in units of $\hbar/\ga$. The blue (green)
solid lines in panel (a) represents the $p_{\Ga}=1/4 ~(3/4)$ curves
in the $\al-T$ plane, while the violet line corresponds to
$p_{\Ga}=1$. The black solid line in panel (b) represents the
$p_K=3/4$ curve in the $\al-T$ plane. The region with no phase band
crossing corresponds to $p_K >3/4$.} \label{fig14} \end{figure}

In this section, we chart out the phase diagram obtained using exact
numerics and the adiabatic-impulse method. At the outset, we note
that all results obtained using exact numerics concur almost exactly
for all $\om/\ga \le 1$ with those obtained from
the adiabatic-impulse approximation discussed in Sec.\ \ref{secadimp}.
One of the central results of our work is the change in topology of
the driven system at $t=T/3$ and $2T/3$ as shown in Figs.\
\ref{fig9}, \ref{fig10}, \ref{fig11}, \ref{fig12}, and \ref{fig13}.
Moreover, we also provide an analysis of the phase diagram of the system at
$t=T$, and we show in Fig.\ \ref{fig14} the contributions to this phase
diagram from different high-symmetry points in the graphene Brillouin zone.
In what follows we list the salient features of the
low-frequency phase diagram of irradiated graphene.

For the computation of such phase diagrams, the standard method followed
in the literature involves two widely followed procedures. The first
involves putting the time-dependent graphene Hamiltonian on a
lattice which is periodic (spatially) along one direction (taken to
be $y$ here) and has an edge (either zigzag or armchair) along the
other direction, $x$.
One then calculates the evolution operator numerically for such a system
and numerically diagonalizes it to obtain the phase bands
\cite{top1,top2,oka1,gil1}. It is well-known \cite{top1,top2,gil1}
that a change in the bulk Chern band would involve a change in the
number of edge states of the phase bands (or equivalently the
Floquet Hamiltonian if we focus on $t=T$). The second method
involves a direct numerical computation of the bulk Chern number in
each of the phases separated by a Floquet topological transition;
the phase diagram can then be charted out by computing the
change in the Chern number across the transition induced by phase band
crossings \cite{oka1,suzuki1}. The computation of the bulk Chern
number involves an integration of the Berry curvature $\vec
{\mathcal B}(k_x,k_y)= \nabla_{\vec k} \times \vec {\mathcal A}(k_x,
k_y)$ over the Brillouin zone, where the Berry potential $\vec {\mathcal
A} = (A_x,A_y)$ can be expressed in terms of the system wave function
$|\psi\rangle$ as ${\mathcal A}_i = \partial_{k_i} |\psi(k_x, k_y;
t_0\rangle$, where the Chern number is computed at $t=t_0$ . This is
typically done by dividing the Brillouin zone into a mesh and summing
up the integrand over all the points on the mesh \cite{suzuki1}. The choice
of the mesh size is of key importance in this procedure and the optimal
choice depends, among other things, on the drive frequency.

It turns out that both these methods lead to significant
computational difficulties at low drive frequencies due to the following
reasons. First, we note that as the frequency is lowered, the Berry
curvature becomes an increasingly rugged function of $\vec k$ and
develops several sharp and near-singular features at different points in
the Brillouin zone whose locations depend on both the drive
frequency and parameters of the system Hamiltonian. Thus a numerical
computation of the integral of ${\mathcal B}$ over the Brillouin
zone which yields the Chern number requires a mesh size which
decreases rapidly with increasing $T$; we find that this makes it
almost impossible to reliably compute this integral for
$\om/\ga \le 1$. Second, for the existence of Floquet edge states
with $E=0$ and $\pi/T$, one needs to have a finite gap for the bulk
Floquet states around this point since otherwise the edge and the
bulk states may hybridize. Such a gap is known to decrease with
decreasing $\om$; also, the number of edge states in the gap
around $E=0, \pi/T$ increases with decreasing $\om$.
Consequently, the determination of a change in the number of edge
states across the transition becomes difficult at low
drive frequencies. Thus, in what follows, we chart out the phase
diagram by noting the transition points where the phase bands cross
at low $T$; we have checked numerically for a few representative
points in this phase diagram that the expected changes in the Chern number
and the number of edge states do indeed occur across such crossing points.

The phase diagram at $t=T/3$ showing such a topology change of the
phase bands is shown in Fig.\ \ref{fig9}. We numerically find that
the entire contribution to this phase diagram occurs from the $\Ga$
point which is in accordance with the prediction from the
adiabatic-impulse theory. We have checked that no crossings occur
for $t=T/3$ at the $K$, $M$, and $X$ points. The change in the Chern
number occurs along the dotted line and the corresponding phase band
crossing always occur through the zone center ($\cos \phi =1$);
these are also in accordance with the predictions of the
adiabatic-impulse method. The computation of the change in Chern
number at a representative point $\al=2.4$ and $T=11.2$ is shown in
Fig.\ \ref{fig10}. Figure \ref{fig10} (a) shows the point where the
phase bands cross through the zone edge. The corresponding Chern
number changes from $0$ to $-1$ at this point as shown in Fig.\
\ref{fig10} (b). Figures \ref{fig10} (c) and (d) show that number of
edge state changes by $-1$ by tabulating their numbers before and
after the crossing. These computations were carried out by standard
numerical methods following Refs.~\onlinecite{top1},
\onlinecite{top2}, \onlinecite{oka1} and \onlinecite{suzuki1} as
discussed earlier in this section. We have checked that for a few
representative points with $T/(2\pi) \le 3$ that such a change in
the Chern number and the number of edge states is consistent with
the position of the phase band crossings in Fig.\ \ref{fig9}.
However, for $T/(2 \pi) \ge 3$, these computations become
numerically difficult for the reasons mentioned above. Finally, we
note that no phase bands crossings occur at $\om/\ga \ge 1$ (or $T
\le 2 \pi$) within the range of $\al$ considered here indicating
that no topology change occurs at high frequencies at $t=T/3$ within
this range. However, at higher $\al$, it is possible that such a
topology change may occur at $t=T/3$ at higher $\om$. We note that
near $\al = 4\pi/(3\sqrt{3}) \simeq 2.418$, all the lines in the
phase diagram show sharp bends; this feature originates from the
fact at this value of $\al$ and all values of $T$, there are
unavoided crossings of the instantaneous energy levels (i.e., the
ground and excited energy levels become exactly degenerate) at
$t=T/12$ and $t=T/4$ which lead to $p_{\Ga} \simeq 1$. To see this
more clearly, we compare the numerical phase-diagram with that
obtained using the adiabatic-impulse method in Fig.\ \ref{fig11}. We
find that there is a near exact match for all values of $\al$ and
$T$ except near $\al \simeq 2.4$; near this line, $p_{\Ga}=1$ and
the adiabatic-impulse method predicts phase band crossings for all
values of $T$ as can be seen from Eq.\ \eqref{pbandgamma1}. This
leads to a vertical line in the $\al-T$ plane as shown in Fig.\
\ref{fig11}. However, exact numerics shows crossings at isolated
points on this line; the density of such points on the line rapidly
increases with $T$ showing that the two results are expected to
match at high $T$. This discrepancy between the results of the
adiabatic-impulse method and exact numerics can be understood to be
the result of unavoided crossings of energy levels. In such a
situation, the two instantaneous energy bands come close to each
other for a wide range of values of $t/T$ around the crossing point.
Thus a transition from the ground state to the excited state can
occur over a wide range of $t/T$ during the evolution. This
invalidates the assumption of a narrow impulse region, unless $T$ is
small.

Next, the phase diagram for $t=2T/3$ is shown in Fig.\ \ref{fig12}.
Here, in contrast to the phase diagram for $t=T/3$, phase band
crossings can occur both through the zone center (black dots) and
zone edges (red dots). The crossings through the zone centers are
same as those seen in Fig.\ \ref{fig9} since by symmetry, the phase
picked between $t=0$ and $T/3$ is same as that between $t=T/3$ and
$2T/3$ at the $\Ga$ point. Thus all phase band crossings for which
$\cos (\phi(T/3))=1$ repeat themselves at $t=2T/3$. In addition, for
$t=2T/3$, phase band crossings may also occur at the zone edge
($\cos(\phi)=-1$). We find that these crossings enclose some closed
regions in the $\al-T$ plane. Each of these regions terminates at
some specific maximal values of $\al=\al_{\rm max}(T)$ in the
$\al-T$ plane. The adiabatic-impulse method predicts that
$p_{\Ga}(\al_{\rm max}(T),T)=1/2$ which matches quite well with the
numerical values of $p_{\Ga}$ at these points as can be seen by
comparing the position of the $p_{\Ga}=1/2$ line to these points
(Fig.\ \ref{fig12}). The changes in the Chern number and the number
of edge states across the transition is shown across a
representative crossing point in Fig.\ \ref{fig13}. In contrast to
the case of the phase diagram at $t=T/3$, topology change occurs
here at high frequencies ($\om/\ga \gtrsim 1$) for the range of
$\al$ considered through the zone edges, as can be seen from Fig.\
\ref{fig12}. 

Next, we consider the phase diagram at $t=T$. Here phase band
crossings occur at several points on the graphene Brillouin zone. We
find, in accordance with the results of the adiabatic-impulse
analysis, that the number of such crossings is large at high
symmetry points. To see this clearly, we plot the phase band
crossings at $t=T$ at the $\Ga$, $K$, $M$ and $X$ points in Fig.\
\ref{fig14}. The plot in Fig.\ \ref{fig14} (a) shows the phase band
crossings that occur at the $\Ga$ point. The closed black and red
dotted regions in the plot correspond to $\phi(T/3)=\pi/3$ (Eq.\
\eqref{gammacrosscond3}) and $\phi(T/3)=2\pi/3$ (Eq.\
\eqref{gammacrosscond4}) respectively.
 As
discussed in Sec.\ \ref{secphdsym}, the adiabatic-impulse method
predicts that such regions should terminate, for a given $T$, at
$\al= \al_{\rm max}(T)$ whose positions should coincide with
$p_{\Ga}=1/4 ~(3/4)$ for regions corresponding to $\phi(T/3)=2\pi/3
~(\pi/3)$. This prediction of the adiabatic-impulse method matches
exact numerics quite well as can be seen by the positions of the
$p_{\Ga}=1/4 ~(3/4)$ curves (green (blue) lines) in Fig.\
\ref{fig14} (a). In Fig.\ \ref{fig14} (b), we show similar crossings
at the $K$ point. The closed regions in the $\al-T$ plane for
crossing through both the zone center (black dots) and the zone edge
(red dots) terminate at some maximal values of $\al = \al_{\rm
max}(T)$ for a given $T$. The positions of these points are
predicted to coincide with the $p_K=3/4$ line in the $\al-T$ plane
(blue line in Fig.\ \ref{fig14} (b)) by the adiabatic-impulse
theory. We find that this prediction matches exact numerics quite
well at high $T$; moreover we find there are no crossings in the
region in $\al-T$ plane for which $p_K
> 3/4$ which is also in accordance with the prediction of the
adiabatic-impulse method. Next, in Fig.\ \ref{fig14} (c), we plot
the crossings at the $M$ point. Here we find that the crossings do
not constitute closed regions or do not terminate along some
specific lines; however, we have checked that their positions for
$T/(2 \pi)>1$ coincides with the predictions of the
adiabatic-impulse method. We also find numerically that there is a
clear demarcation between the zone center and zone edge crossings;
the latter happens only at high $T$ while the former occurs at small
$T$. Finally, in Fig.\ \ref{fig14} (d), we plot the phase band
crossings at the $X$ point which is a lower-symmetry point compared
to $\Ga$, $K$, and $M$. Here we find that phase band crossings occur
at isolated points in the $\al-T$ plane; moreover, the number of
such crossings is much smaller than those at the high symmetry
points. This is in accordance with the general expectation that
phase band crossings are more likely to happen at high symmetry
points in the graphene Brillouin zone.

\begin{figure}[t!]
\begin{center}
\includegraphics[width=\columnwidth]{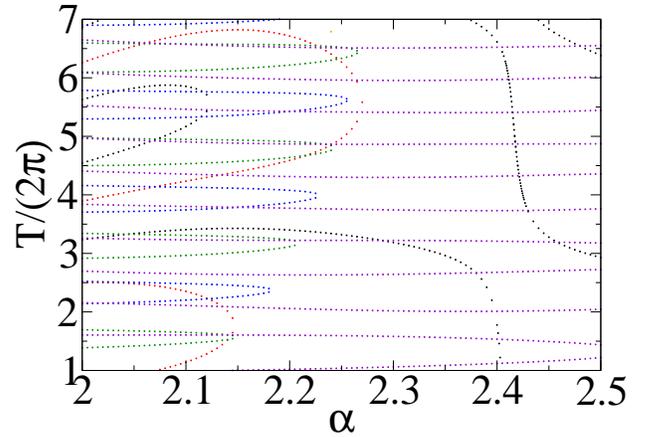}
\end{center}
\caption{Phase diagram at $t=T$ in the $\al-T$ plane. The
contributions from the $\Ga$ point for crossings through the zone
center (edge) are given by black (red) dots. The corresponding
crossings from $K$ points are given by green (zone center) and blue
(zone edge) dots and those from $M$ points correspond to violet
(zone center) and orange (zone edge) dots. See text for details.}
\label{fig14p}
\end{figure}

Finally, we present the phase diagram for $t=T$ in a region in the
$\alpha-T$ plane shown in Fig.\ \ref{fig14p}. We note that for this
region of the $\alpha-T$ plane, phase band crossings occur at $\Ga$,
$K$, and $M$ points; there are no crossings from the $X$ point. We
find that the phase diagram constitutes closed regions each of which
is expected to have a definite Chern number. Each of the lines in
the diagram corresponds to a change in Chern number by $\Delta C=\pm
1$; when two or more such lines coalesce, it indicates a possible
Chern number change by $\pm 2, \pm 3 ...$. However, a detailed
calculation of these Chern at high values of $T/2\pi$ in the phase
diagram turns out to be numerically difficult for reasons discussed
earlier in the text and we have not attempted it here. We also note
that the phase diagram may change due to possible non-generic phase
band crossings at other low-symmetry points of the graphene
Brillouin zone. Such crossings are expected to be very few in number
for reasons discussed earlier in Sec.\ \ref{secphdsym} and our
numerics have not found any for the graphene Brillouin zone points
that we have checked.  However, we note that it is not numerically
possible to completely ascertain their non-existence for all points
in the graphene Brillouin zone and for the entire  $\alpha-T$ plane.

\section{1D $XY$ model in a transverse field}
\label{sec1dxy}

We now turn to the one-dimensional spin-1/2 model with $XY$
nearest-neighbor interactions and a magnetic field $\mu$ applied
along the $\hat z$ direction. The Hamiltonian for a system with $N$
sites and open boundary conditions is given by \beq H ~=~ -
~\sum_{n=1}^{N-1} ~[J_x \si_n^x \si_{n+1}^x ~+~ J_y \si_n^y
\si_{n+1}^y] ~-~ \mu ~\sum_{n=1}^N \si_n^z, \label{ham1Ds} \eeq
where the $\si_n^a$'s denote Pauli spin matrices at site $n$. This
can be mapped to a system of spinless fermions by the Jordan-Wigner
transformation; the fermionic Hamiltonian is given by
\cite{rev3,diva1}\bea H &=& \sum_{n=1}^{N-1} ~[ \ga (f_n^\da f_{n+1}
+ H.c.) ~+~ \De (f_n
f_{n+1} + H.c.)] \non \\
&& -~ \mu ~\sum_{n=1}^N ~(2 f_n^\da f_n - 1), \label{ham1Df} \eea
where $\ga = J_x + J_y$, $\De = J_y - J_x$, and the fermionic
operators satisfy the usual anticommutation relations $\{ f_m, f_n
\} = \{ f_m^\da, f_n^\da \} = 0$ and $\{ f_m, f_n^\da \} =
\de_{mn}$. In this fermionic language, $\ga$ is a nearest-neighbor
hopping amplitude, $\De$ is a $p$-wave superconducting pairing, and
$2\mu$ is as a chemical potential. It is convenient to define two
Majorana fermion operators at each site, $a_n$ and $b_n$, as \beq
f_n ~=~ \frac{1}{2} (a_n + i b_n) ~~~{\rm and}~~~ f_n^\da ~=~
\frac{1}{2} (a_n - i b_n). \eeq These operators are Hermitian and
satisfy the relations $\{ a_m, a_n \} = \{ b_m, b_n \} = 2 \de_{mn}$
and $\{ a_m, b_n \} = 0$. In terms of these operators, the
Hamiltonian takes the form
\bea H &=& - \frac{i}{2} ~\sum_{n=1}^{N-1} ~[(\ga + \De) a_n b_{n+1} ~+~
(\ga - \De) a_{n+1} b_n] \non \\
&& - i \mu ~\sum_{n=1}^N ~a_n b_n. \label{ham1Dm} \eea
Note that the Hamiltonian is invariant under a parity transformation which
reflects the system about its midpoint, namely, $a_n \to b_{N+1-n}$
and $b_n \to - a_{N+1-n}$.

For a system with $N$ sites and periodic boundary conditions, we can
write the Hamiltonian in momentum space as follows. Defining the
Fourier transform as $f_k = \frac{1}{N} \sum_{n=1} f_n e^{-ikn}$, we
find that Eq.~\eqref{ham1Df} can be rewritten as
\bea H &=& \sum_{0 \le k \le \pi} ~\left( \begin{array}{cc} f_k^\da & f_{-k}
\end{array} \right) H_k ~\left( \begin{array}{c}
f_k \\
f_{-k}^\da \end{array} \right), \non \\
H_k &=& 2(\ga \cos k ~-~ \mu) ~\tau_z ~+~ 2 \De \sin k ~\tau_y, \eea
where the $\tau_a$'s denote Pauli pseudospin matrices in the
particle-hole space. It is convenient to do a unitary transformation
by rotating around $\hat z$ to convert $\tau_y \to \tau_x$ so that
the Hamiltonian in $k$ space is given by \beq H_k ~=~ 2(\ga \cos k
~-~ \mu) ~\tau_z ~+~ 2 \De \sin k ~\tau_x. \label{hamk} \eeq Since
this $H_k$ is a real and symmetric matrix, it will be easier to
derive the symmetry properties of the corresponding Floquet operator
$U_k (t)$.

\subsection{Two-rate protocol}
\label{sectworate}

In what follows, we shall consider the driving of the chemical potential and
superconducting pairing with two different frequencies $\om$ and $r \om$, where
$r$ is an integer, so that
\beq \mu ~=~ A ~\cos (\om t) ~~~{\rm and}~~~ \De ~=~ B \cos (r \om t).
\label{dr2} \eeq
This allows us to write
\bea H_k (t) &=& f_3 (k,t) ~\tau_z ~+~ f_1 (k,t) ~\tau_x, \label{hamk2} \\
f_3(k,t) &=& 2[\ga \cos k ~- A \cos (\om t)], \non \\
f_1(k,t) &=& 2 B \cos (r \om t) \sin k. \non \eea

We now consider the time evolution operator \beq U_k (t) ~=~ {\cal
T}_t ~\exp [-i \int_0^t dt' H_k (t')]. \eeq We will be particularly
interested in the conditions under which $U_k (t)$ will be equal to
$\pm I$, giving a phase band crossing.

A simple way of getting $U_k (t) = \pm I$ is to set $k=k_0$ where
$k_0 =0$ or $\pi$; then $\sin k_0 = 0$ and $f_1 (k_0, t) = 0$ for
all $t$. This gives \beq U_{k_0} (t) ~=~ \exp [-i \tau_z \int_0^t
dt' f_3 (k_0,t') dt']. \label{ut8} \eeq The phase band crossing
condition then takes the form \beq - ~\int_0^t dt' f_3 (k_0,t') dt'
~=~ n \pi, \label{npi} \eeq where $n$ is an integer. For Eq.\
\eqref{hamk2}, we can analytically find if Eq.~\eqref{npi} has
solutions for any value of $t$. Such phase band crossings which
occur at the edge or center of the 1D Brillouin zone has been
studied in details in Ref.\ \onlinecite{bm1}; we shall not discuss
them further here.

In addition to phase band crossings through the zone edge or center
discussed above, we find that two-rate protocols may lead to
additional phase band crossings at $k \ne 0, \pi$. For
Eq.~\eqref{hamk2}, there exists such a phase band crossing at
$k=\pi/2$. For odd integer $r$, we see that \beq H_{\pi/2} (T/2 - t)
~=~ -~ H_{\pi/2} (t). \label{hamtby2} \eeq Using Eq.~\eqref{hamtby2}
and the form of the time evolution operator given in
Eq.~\eqref{ut2}, we then see that \beq [U_{\pi/2} (T/2)]^{-1} ~=~
U_{\pi/2} (T/2). \eeq This implies that $U_{\pi/2} (T/2) = \pm I$.
Thus Eq.~\eqref{hamk2} gives a phase band crossing at $k=\pi/2$ and
$t=T/2$ for any value of $\ga, ~A, ~B$ and $T$, if $r$ is an odd
integer. Further, we have \beq H_{\pi/2} (T/2 + t) ~=~ -~ H_{\pi/2}
(t) \label{hamtby2p} \eeq if $r$ is an odd integer. This implies
that \beq U_{\pi/2} (T) ~=~ [U_{\pi/2} (T/2)]^T ~U_{\pi/2} (T/2).
\eeq This, combined with $U_{\pi/2} (T/2) = \pm I$ implies that
$U_{\pi/2} (T) = I$ for any value of $\ga, ~A, ~B$ and $T$.

Assuming $r$ to be an integer, we see from Eq.~\eqref{hamk2}
\beq H_k (T - t) = H_k (t) \label{hamtmt} \eeq
for any value of $k$. Following the arguments
presented in Eqs.~(\ref{ut2}-\ref{ut3}) we see that $U(T)$ must
again be of the form given in Eq.~\eqref{ut3}. Hence we expect that
we should be able to find a phase band crossing at $t=T$ by varying
two parameters, such as $\om$ and $k$.

\begin{figure}
\includegraphics[height=57mm,width=\linewidth]{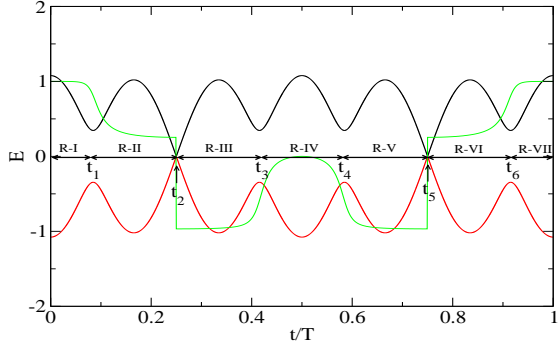}
\caption{Instantaneous energies of the ground (lower red curve) and
excited (upper black curve) states and $\eta$ (green curve) vs $t/T$
for $k=\pi/2$, $A=0.4$, $r=3$, and $\om=0.1$.} \label{fig15}
\end{figure}

Next, we analyze such phase band crossings for Hamiltonian given by
Eq.~\eqref{hamk2}. To this end, we first study the eigenvalues of
the instantaneous Hamiltonian $H_k (t)$ as a function of $t/T$ for
$k=\pi/2$. This is shown in Fig.~\ref{fig15}. There are seven
regions, denoted by ${\rm R-i, ~i= I, II, III, \cdots, VII}$, where
there is a substantial gap between the ground and excited state
energies and the evolution of the system is adiabatic. Any two such
regions R-i and R-\{i+1\} are separated by an avoided crossing point
$t_i$ where the gap of the instantaneous Hamiltonian has a minimum.
There are six such times, $\{t_i, i=1, \cdots, 6\}$, and there is a small
region around each of these times where we can use the impulse approximation
to calculate the Landau-Zener transition probability. The analysis
follows the same route as charted out in Sec.\ \ref{secadimp} and the
Appendix. In what follows, we shall just present the final results obtained by
calculating the phase band $\cos(\phi(t))$ for $k=\pi/2$ and $t = T/2$.

\begin{figure}
\includegraphics[height=57mm,width=\linewidth]{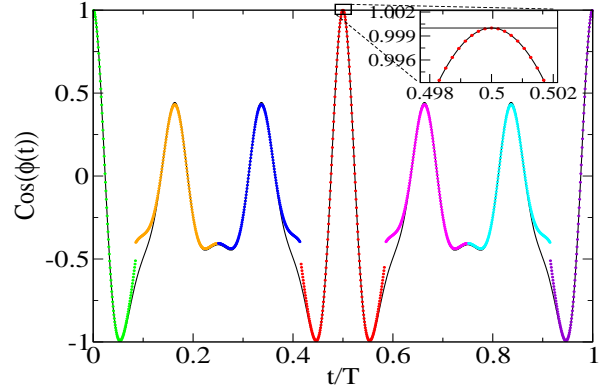}
\caption{Comparison of numerical and adiabatic-impulse results for $k=\pi/2$,
$A=0.4$, $r=3$, and $\om=0.1$.} \label{fig16} \end{figure}

We first show a comparison in Fig.~\ref{fig16} between the
numerically calculated plot of $\cos (\phi (t))$ versus $t/T$ and
the result obtained by the adiabatic-impulse method for $k=\pi/2$,
$A=0.4$, $r=3$, and $\om = 0.1$. We see that the match is excellent
except in the small impulse regions. In particular, the agreement
between the numerical and adiabatic-impulse results is found to be
extremely good in region-IV around the phase band crossing at
$t=T/2$. In fact, using the different symmetries which are clearly
visible in Fig.~\ref{fig15} (such as
$\xi_k(t_2,t_1)=\xi_k(t_3,t_2)$, $p_{1k} =p_{3k}$ and
$\tilde{\phi}_{1k} =\tilde{\phi}_{3k}$ for $k=\pi/2$), one can
obtain a simple expression for $\cos (\phi (t=T/2))$, \beq
\cos(\phi(T/2))~=~\cos[\xi(T/2,t_3)-\xi(t_1,0)], \eeq where
\begin{widetext}
\bea \xi_k(t_f,t_i) &=& \int_{t_i}^{t_f} dt [(A\cos(\om t)-\cos k)^2 ~+~
\cos^2(r \om t) \sin^2 k], \non \\
p_{ik} &=& \exp[-2\pi\de_{ik}], ~~~~ \de_{ik} =|\be_{ik}|^2/(2|\al_{ik}|),
\non \\
\al_{ik} &=& \sqrt{A^2 \om^2 \sin^2 (\om t) ~+~ r^2 \om^2 \sin^2 (r \om t_i)
\sin^2 k}, \non \\
\be_{ik} &=& \sqrt{[A \cos (\om t_i) - A \om \sin (\om t_i) t'_i]^2 ~+~
[\cos (r \om t_i) \sin k - r \om \sin (r \om t_i) t'_i \sin k]^2}, \non \\
t'_i &=& [A^2 \om \sin (\om t_i) \cos (\om t_i) ~+~ r \om \sin (r \om t_i)
\cos (r \om t_i) \sin k]/(\al_{ik})^2, \non \\
\tilde{\phi}_{ik} &=& - ~3\pi/4 ~+~ \de_{ik} [\ln (\de_{ik})-1] ~+~
{\rm Arg} \Ga(1-i\de_{ik}). \eea
\end{widetext}
Here we have used the fact that at $t=T/2$ we have $\eta=0$ and
$\xi(T/2,t_3)=\xi(t_1,0)$. Hence $\cos(\phi(T/2))=1$ at all
frequencies. Thus the adiabatic-impulse method becomes exact at this
point and is valid at all frequencies. This is corroborated by
checking how the match between the numerical and adiabatic-impulse
results changes as the frequency is increased as shown in
Fig.~\ref{fig17}. We see that at higher frequencies there is an
increasing deviation between the two results as we go away from the
phase band crossing point $t=T/2$ on both sides, as is expected;
however the match at the crossing point remains exact. Thus we find
that both adiabatic-impulse and general symmetry arguments predict a
line of phase band crossings at $t=T/2$ and $k=\pi/2$ for any $T$ in
this model; we have checked that this prediction agrees with exact
numerics. We note that any such phase band crossings at $t=T/2$ also
lead to a crossing at $t=T$ since
$U_{\pi/2}(T/2,0)=U_{\pi/2}(T,T/2)$ which follows from the symmetry
condition $H_k(T-t)=H_k(t)$. However the analysis of such crossings
follows the same line as our earlier analysis and we do not repeat
it here. Instead, in Sec.\ \ref{endmodes}, we concentrate on the
structure of the end modes of such a driven Hamiltonian on a chain
with a finite length. Our numerics with the model also suggests that
there are no other generic crossings for any $k \ne 0,\pi/2,\pi$ for
odd integer $r$; this is consistent with the symmetry analysis
carried out earlier in this section.

\begin{widetext}
\begin{center}
\begin{figure}
\subfigure[]{\includegraphics[height=60mm,width=.46\linewidth]{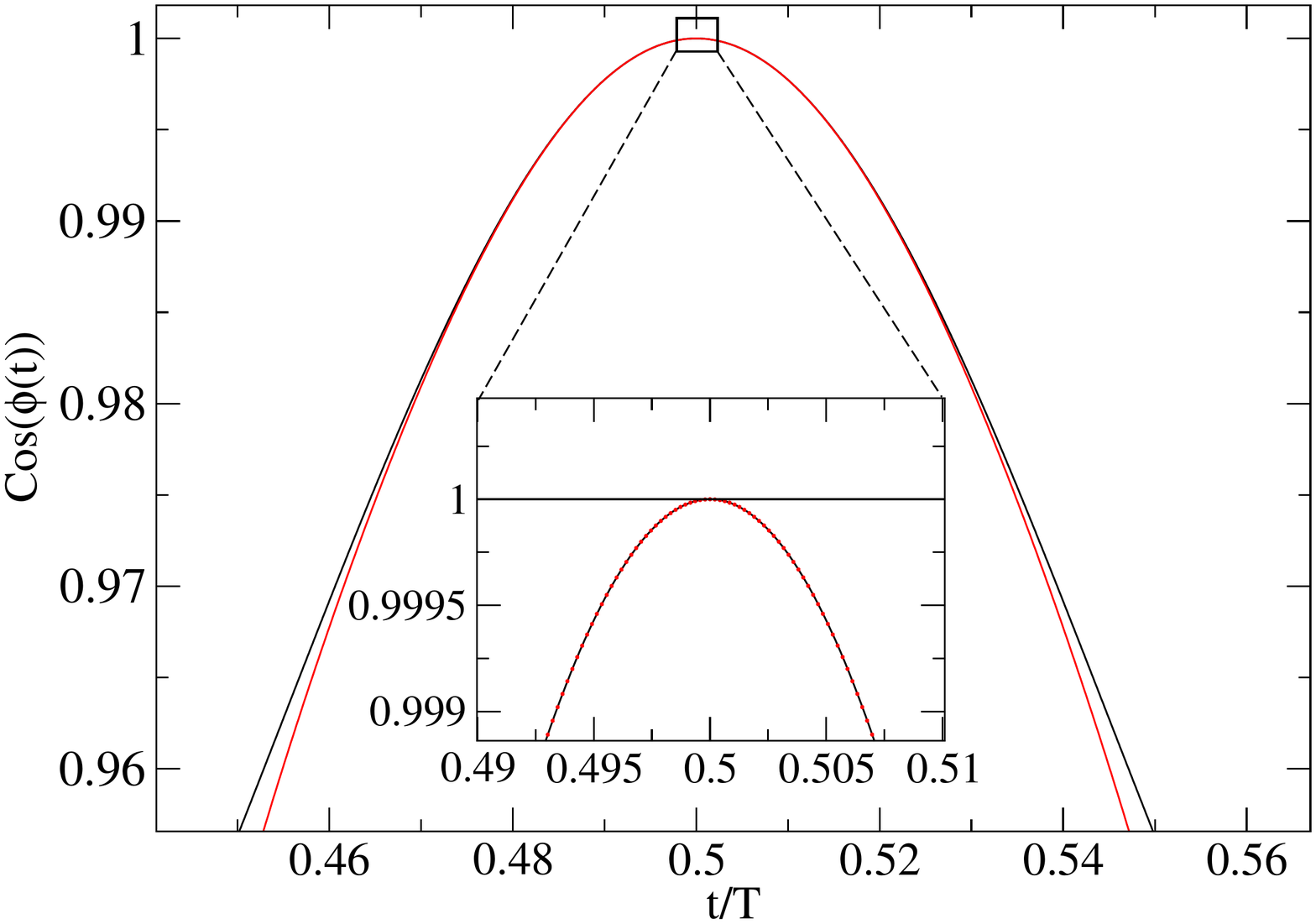}}
\subfigure[]{\includegraphics[height=60mm,width=.45\linewidth]{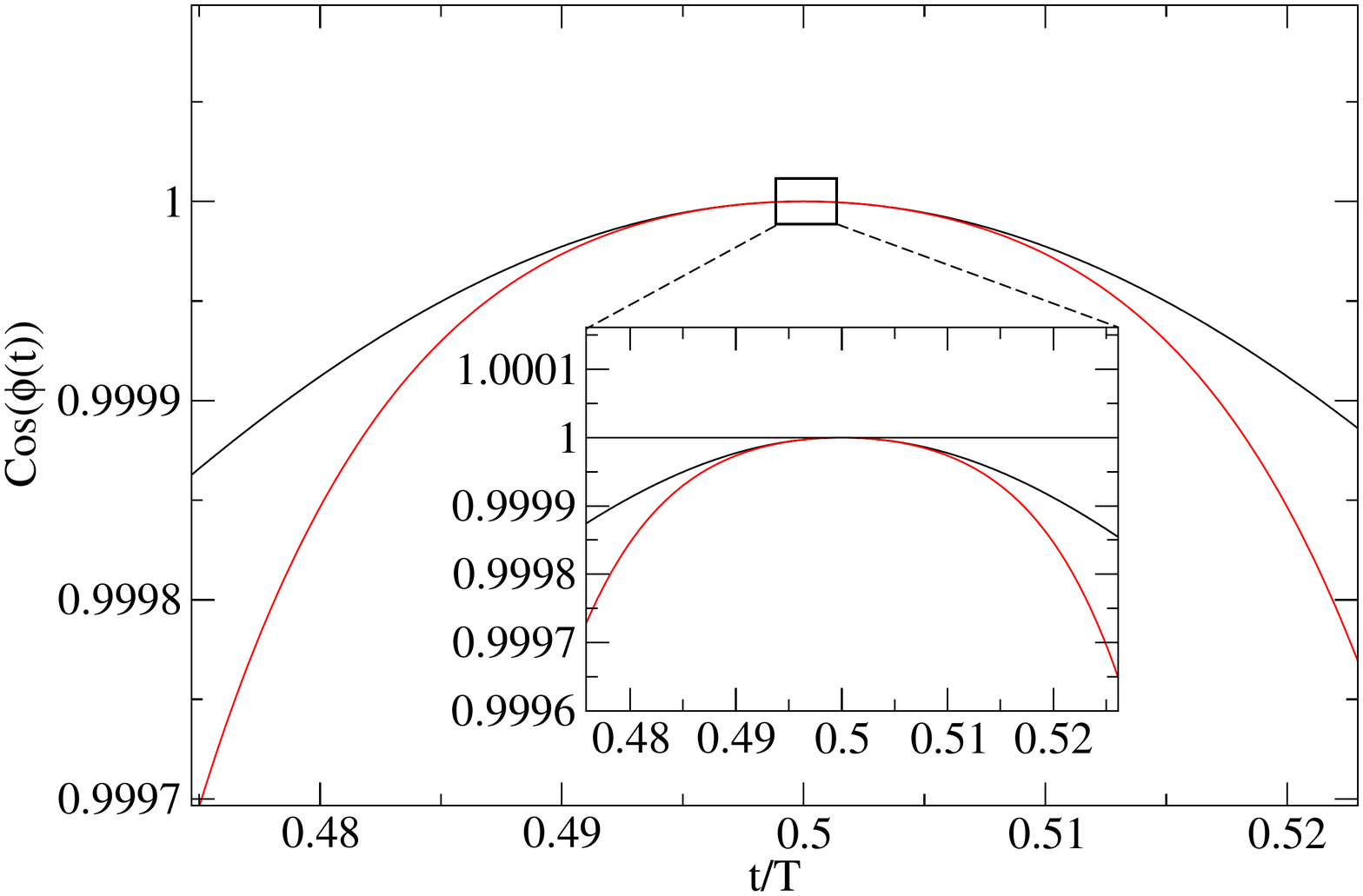}}
\caption{Comparison of numerical and adiabatic-impulse results at higher
frequencies, (a) $\om=1$ (top) and (b) $\om=10$ (bottom), for $k=\pi/2$,
$A=0.4$ and $r=3$.} \label{fig17} \end{figure}
\end{center}
\end{widetext}

\subsection{End modes}
\label{endmodes}

A system with $N$ sites and open boundary
conditions sometimes has end modes depending on the different
parameters. An end mode is an eigenvector of the time evolution
operator $U(t)$ whose wave function is localized near one of the two
ends of the system. Due to the parity symmetry of
Eq.~\eqref{ham1Dm}, end modes always occur in pairs with one mode at
each end; the two modes do not hybridize if the system size is much
larger than the decay length of each mode. We note that an end mode
can appear or disappear when we go across a phase band crossing
occurring at, say, $t=t_0$; the way this happens is that the decay
length of the end mode diverges as $t \to t_0$ so that the end mode
merges with the bulk modes at $t=t_0$. We will consider below some
properties of end modes when we are away from a phase band crossing.

To understand the nature of the end modes, it is useful to write the
Hamiltonian $H$ and $U$ in terms of Majorana operators
\cite{abhiskar1}. We define a $2N$-dimensional column $c$ whose
entries $c_n$ are given by $a_1, b_1, a_2, b_2, \cdots, a_N, b_N$.
The Hamiltonian in Eq.~\eqref{ham1Dm} can then be written in the form
\beq H ~=~ \frac{i}{2} ~\sum_{m,n=1}^{2N} ~c_m M_{mn} c_n, \label{hamcMc} \eeq
where $M$ is a real antisymmetric matrix.

Next, we allow $H$ and $M$ to vary with time. The Heisenberg
operators $c_n (t)$ satisfy the equations \beq \frac{dc_n (t)}{dt}
~=~ i [ H(t), c_n (t)]. \eeq Eq.~\eqref{hamcMc} then implies that
\beq \frac{dc_m (t)}{dt} ~=~ 2 \sum_{n=1}^{2N} ~M_{mn} (t) c_n (t).
\label{eom} \eeq In terms of the column $c(t)$ and the matrix
$M(t)$, the solution of Eq.~\eqref{eom} can be written as
\bea c(t) &=& U(t) ~c(0), \non \\
{\rm where} ~~~U(t) &=& {\cal T}_t \exp [~2 \int_0^t dt' M(t')]. \eea
We thus see that $U(t,0)$ is a real and unitary matrix; hence it is also
orthogonal.

We now look at the eigenvectors and eigenvalues of $U(t)$. If $x$ is
an eigenvector of $U(t)$ with eigenvalue $e^{i\theta}$, the fact
that $U(t)$ is real implies that $x^*$ is an eigenvector of $U(t)$
with eigenvalue $e^{-i\theta}$. This implies that for eigenvalues
equal to $\pm 1$, $x$ and $x^*$ are degenerate; hence the eigenvectors can
be chosen to be real by taking the combinations $x+x^*$ and $i (x-x^*)$. In
particular, end modes with eigenvalues equal to $\pm 1$ will have real
eigenvectors; such modes with real wave functions are called Majorana end
modes \cite{abhiskar1}, in analogy with the Majorana end modes of
time-independent Hamiltonians with time-reversal symmetry~\cite{degottardi}.

Sometimes we find end modes for which the eigenvalues of $U(t)$ are
{\it not} equal to $\pm 1$; these are called anomalous end modes
\cite{saha}. Such modes always occur in pairs at each end of the
system, with the eigenvalues of the pair being complex conjugates of each
other. Further, the wave functions of such modes are necessarily complex.

\begin{widetext}
\begin{center}
\begin{figure}[t!]
\begin{center}
\subfigure[]{\includegraphics[width=2.3in]{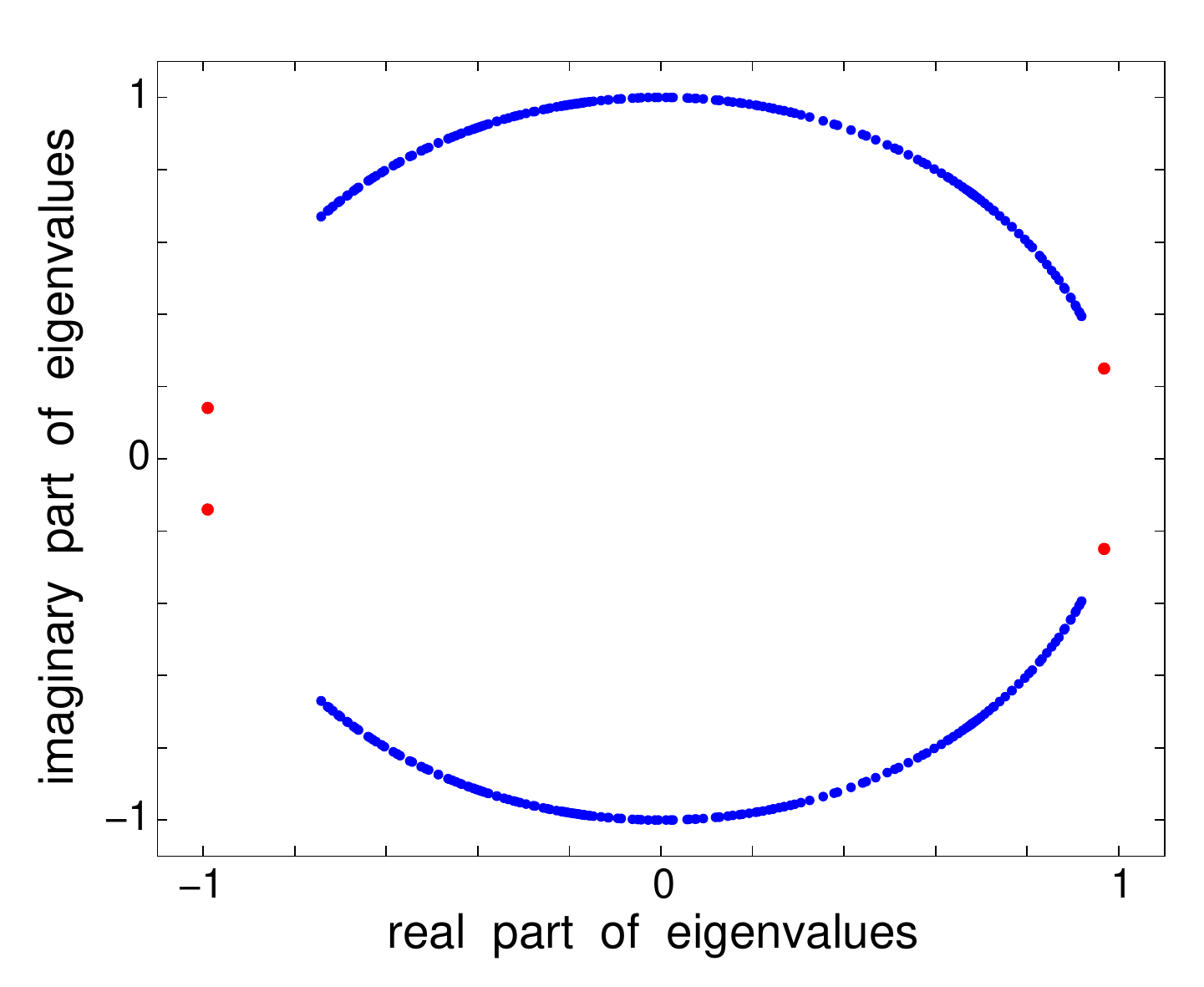}}
\subfigure[]{\includegraphics[width=2.3in]{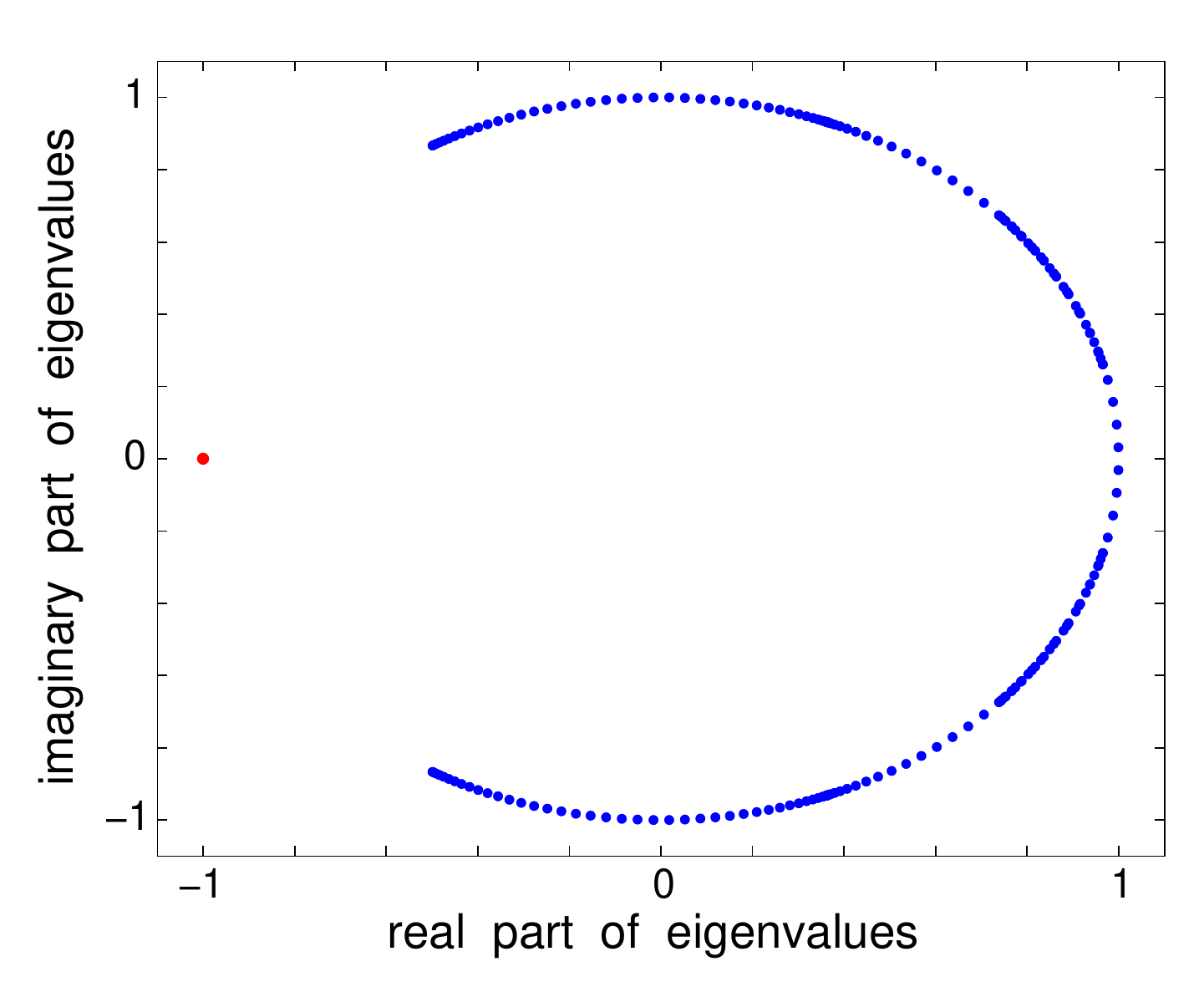}}
\subfigure[]{\includegraphics[width=2.3in]{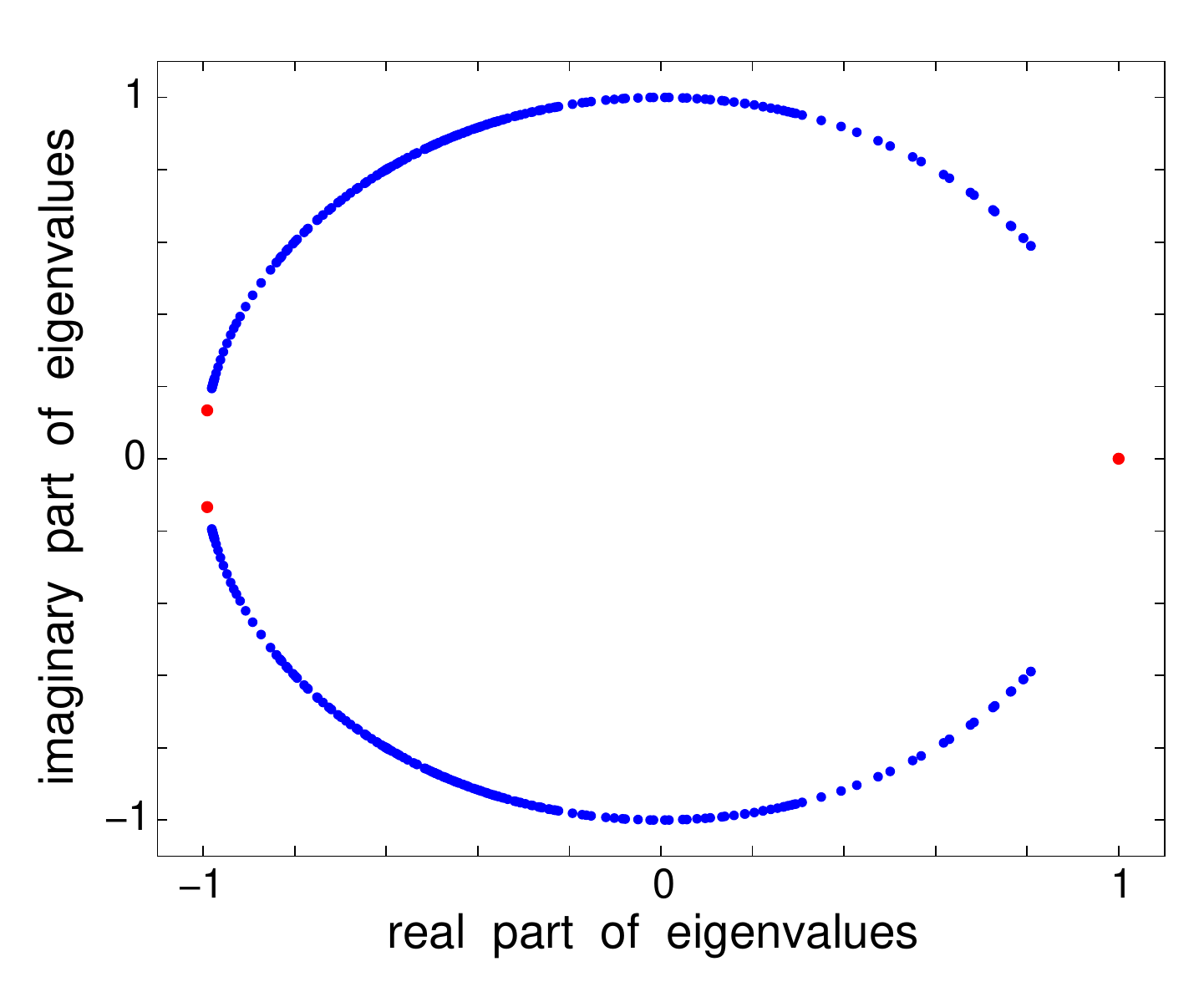}}
\end{center}
\caption[]{Real and imaginary parts of the eigenvalues $e^{i\theta}$ of the
Floquet operator $U(T/2)$ for a 200-site system driven as in Eq.~\eqref{ham1Dm}
and \eqref{dr2}. We have taken $\ga = 1, ~A=0.4, ~B=1, ~r=3, ~\om =0.3$, and
(a) $t/T= 0.45$, (b) $t/T=0.5$, and (c) $t/T = 0.55$. The isolated eigenvalues
(shown as red dots) correspond to end modes, while the continuous arcs of
eigenvalues correspond to bulk modes.} \label{fig:floeig}
\end{figure}
\end{center}
\end{widetext}

Fig.~\ref{fig:floeig} shows the eigenvalues $e^{i\theta}$ of the
Floquet operator $U(T)$ for a 200-site system driven as described in
Eqs.~\eqref{ham1Dm} and \eqref{dr2}. The parameter values are $\ga =
1, ~A=0.4, ~B=1, ~r=3, ~\om = 0.3$, and (a) $t/T= 0.45$, (b)
$t/T=0.5$, and (c) $t/T = 0.55$. In all three cases, a large number
of end modes are present; these modes are visible in the figures as
isolated eigenvalues which are shown as red dots. All these
eigenvalues have at least a two-fold degeneracy due to the parity
symmetry, i.e., there are always a pair of modes lying at opposite
two ends of the system with the same Floquet eigenvalue. In
Fig.~\ref{fig:floeig} (a), we see eight anomalous end modes (each of
the four isolated eigenvalues has a two-fold degeneracy). In
Fig.~\ref{fig:floeig} (b), we see four Majorana end modes with
Floquet eigenvalue equal to $-1$ (this eigenvalue has a four-fold
degeneracy). In Fig.~\ref{fig:floeig} (c), we see two Majorana end
modes with Floquet eigenvalue equal to $+1$ (with a two-fold
degeneracy) and four anomalous end modes (each of the two
eigenvalues has a two-fold degeneracy). In Figs.~\ref{fig:floeig}
(a) and (c), we see that the continuous parts of the Floquet
eigenvalue spectrum (these correspond to the bulk modes) form two
disjoint arcs which are separated by gaps around $e^{i\theta}$ equal
to both $+1$ and $-1$ (i.e., zone center and zone edge). In
Fig.~\ref{fig:floeig} (b), the continuous part forms a single arc
with no gap around $e^{i\theta} = 1$ (zone center); thus there is a
phase band crossing at $t=T/2$ with $U_{k=\pi/2} = + I$. For all the
end modes, we have checked numerically that the Majorana modes with
Floquet eigenvalues equal to $\pm 1$ have purely real wave
functions, while the anomalous modes appear in pairs with complex
conjugate eigenvalues ($e^{\pm i \theta}$) and their wave functions
are complex. Interestingly, we observe that there are no anomalous
end modes at $t/T = 0.5$ where a phase band crossing occurs
(Fig.~\ref{fig:floeig} (b)).

\section{Discussion}
\label{secdiss}

In this work, we have studied a class of driven closed quantum
integrable systems in the presence of either an external radiation
or a two-rate drive protocol at low frequencies. Such systems have
been studied before either in the regime where the frequency of
radiation is high or for models where a single parameter of the
Hamiltonian is driven periodically. Our study therefore provides a
complementary set of results to the existing literature.

For graphene in the presence of external radiation, we study the
change in topology of the system by studying its time evolution
operator $U$. Such studies have been carried out in the literature
earlier in the high-frequency regime where perturbative $1/\om$
expansions work. In contrast, our work addresses this phenomenon in
the low-frequency regime where such perturbative treatments fail. To
obtain an analytical understanding of the phase diagram of graphene
in the low-frequency regime we therefore have used the
adiabatic-impulse approximation by appropriately modifying it for
the present drive protocol. We note that such an approximation
yields results for the phase bands of the system which provides a
near exact match with exact numerics at low frequencies; moreover,
it allows us to provide semi-analytic criteria for the conditions of
the phase band crossings in graphene. Our analysis also shows that
the phase band crossings, leading to change in Chern number of the
phase bands in graphene, are generically expected to occur at the
high-symmetry points in the graphene Brillouin zone such as $\Ga$,
$K$ and $M$ points. We also provide, for each of these points,
analytic criteria for such crossings. Our results indicate the
presence of such crossings at $t=T/3$ and $2T/3$ (apart from those
at $t=T$) indicating the inadequacy of a Floquet Hamiltonian based
analysis. Such crossings at $t=T/3$ and $t=2T/3$ lead to a distinct
phase diagram whose contribution comes solely from the $\Ga$ point
in the graphene Brillouin zone; such crossings are shown not to
occur at high frequencies for a range of radiation amplitude
$\alpha$. This explains why such diagrams can not be obtained by
perturbative techniques which rely on some form of $1/\om$
expansions. In contrast at $t=T$, we find that the phase band
crossings may occur at other high symmetry points such as $K$ and
$M$ in the graphene Brillouin zone; the general phase diagram at
$t=T$ receives contribution from all such points. We emphasize that
several aspects of such phase band crossings can be analytically
understood from an analysis using the adiabatic-impulse method which
we carried out in this work; they lead to semi-analytic conditions
for several aspects of the phase diagram (Eqs.\
(\ref{gammacrosscond1}-\ref{gammacrosscond4}),
\eqref{diraccrosscond}, and \eqref{Mcrosscond}) which match almost
exactly with numerics. Moreover, such an analysis indicates the role
of symmetry of the irradiated graphene Hamiltonian behind such phase
band crossings in the low-frequency regime and predicts exact
semi-analytic conditions (Eqs.\ \ref{finalcond}, \ref{evencos1}, and
\ref{oddcos1}) for phase band crossings for any $t \le T$. Our study
takes note of the difficulty in generic numerical computation of the
Chern number and its change in the low-frequency regime; this also
allows us to point out the benefit determining the position of the
phase band crossings (which can be done reliably for any frequency)
in determining the structure of the graphene phase diagram at low
frequency. We also find that whereas generic phase band crossings
are expected to occur only at high symmetry points, one can not rule
out the presence of accidental crossings at other times and for
other values of $\vec k$; however, the number of such accidental
crossings are expected to be much lower than the generic ones.

For the 1D $XY$ model we have shown that a two-rate drive protocol
leads to additional phase band crossings at $t=T/2$ for $k=\pi/2$.
for any drive frequency when the ratio $r$ of the two drive
frequencies is an odd integer. In contrast to the earlier studied
phase band crossings in this model for a single parameter drive
protocol which occur at $k=0, \pi$, for a two-rate drive, we find
crossings at $k=\pi/2$ and $t=T/2$ which have no counterpart for
single parameter drive protocols. Moreover such crossings at $t=T/2$
lead to additional phase band crossings at $t=T$ for $k=\pi/2$; this
was shown via a symmetry analysis of the model. These phase band
crossings for $k=\pi/2$ occur for any value of $T$ as long as the
ratio of drive frequencies, $r$, is an odd integer. We have also
studied the end modes of such a driven model for a finite a chain
with $N$ sites. We have found, apart from the usual Majorana end
modes, the existence of anomalous end modes~\cite{saha}. No
anomalous end modes are found at the phase band crossing points at
$t=T/2$ where only Majorana modes exist; across this point the
number of end modes change confirming a topological phase
transition.

There are several experiments which may confirm our theoretical results. The
simplest among them would be look for the phase diagrams of irradiated
graphene at $t=T/3$ and $t=2T/3$ at low drive frequencies $\om/\ga
\simeq 0.1$ and for drive amplitudes $2 \le \alpha \le 2.5$; these are shown
in Figs.\ \ref{fig9} and \ref{fig12}. These phase diagrams would be much
simpler to verify experimentally since the only contribution to them comes
from the $\Ga$ point in the graphene Brillouin zone. We note in this
context that angle-resolved photoemission spectroscopy (ARPES) has
already been performed on irradiated graphene leading to a detection
of Chern number change via photoelectron intensity measurements
\cite{exp2,exp3}. In fact such experiments, by a variation of
the intensity of the applied photon, can pick out selective contributions to
the Chern number change from different points in the graphene
Brillouin zone \cite{exp2}. For example, the contribution to the
phase diagram coming from the $K$ and $K'$ points would show up in such
measurements; thus we expect such measurement to reproduce the
phase diagram in Fig.\ \ref{fig14} (b). The presence of chiral edge
modes in such driven systems has also been experimentally verified
recently both in phononic crystals \cite{exp4} and in transport
experiments on topological insulator surfaces \cite{exp5}; a change in their
number across a phase band crossing should also lead to a reconstruction of
the phase diagrams that we provide here provided that such experiments can
be suitably modified and designed for graphene-like systems.

Our study also leads to several open questions which we intend to
study in the future. For example, it would be interesting to study the
properties of transport in irradiated graphene-like systems in the
presence of disorder. An analogous problem has been studied for high
radiation frequencies in Refs.\ \onlinecite{kundu1}; however, given
the complexity of the phase diagram, its low-frequency counterpart
is expected to lead to several and yet unexplored features in the transport.
Further, the effect of a weak interaction in these Dirac systems
which breaks its integrability would be interesting to study; such
studies are clearly numerically difficult in higher dimensions,
and we expect a suitably modified version of the adiabatic-impulse
approximation to shed some light in this matter.

To conclude, we have studied the low-frequency phase diagram of
irradiated graphene and a driven $XY$ model. Our study constitutes an
application of a suitably modified adiabatic-impulse approximation to
address the dynamics of these models. In the low-frequency regime,
for both systems, the analytical results obtained using this
approximation provides a near-exact match with numerics. This allows
us to provide semi-analytic criteria for phase band crossings and
hence a change in the topology of the wave function of such systems.
Our analysis predicts a change in the topology for irradiated graphene
at $t=T/3$ and $2T/3$ and has provided the corresponding phase
diagrams; such diagrams indicate the inadequacy of a Floquet
Hamiltonian based analysis for these systems which can only provide
information about the phase diagram at $t=T$. Our work also shows that such
a change in the topology of the 1D $XY$ model driven using a two-rate
protocol may occur at $t=T/2$ and points out a change in the end mode
structure across this transition for a $XY$ chain with a finite length.
Finally we have suggested several experiments which may test our theory.

\appendix

\section{Computation of phase bands for $n$ avoided crossings}

In this appendix, we provide expressions for the eigenvalues of the
unitary evolution operators $U_{\vec k}(t_f,0)$ for $n$ avoided
level crossings, where $t_f$ denotes the time at which the phase
bands need to be computed. To this end, we first construct $U_{\vec
k}^{\rm ad}(t_f,0)$ for each of the adiabatic regimes between two
avoided level crossings using Eq.\ \eqref{uadrel1}, and then relate
it to $U_{\vec k}(t_f,0)$ using Eq.\ \eqref{etadef}.

In the adiabatic region prior to the first avoided crossing, the
wave function merely gathers some kinematic phase. Thus we have,
from Eq.\ \eqref{uadrel1} for $t_f \le t_{1 \vec k}$, \cite{nori1}
\bea \vec c^{(1)}_{\vec k}(t_f) &=& e^{-i \tau_3 \xi_{1\vec k}(t_f)}
\left(
\begin{array}{c} 1 \\ 0
\end{array} \right) = \left(\begin{array}{c} e^{-i \xi_{1\vec k}(t_f)} \\ 0
\end{array} \right), \non \\
\xi_{1\vec k}(t_f) &=& \int_0^{t_f} E_{\vec k}(t') dt',
\label{adregion1} \eea
where $E_{\vec k}(t')$ is given by Eq.\ \eqref{eigen1}. Using Eqs.\
\eqref{adregion1} and \eqref{etadef}, we obtain
\bea [U^{(1)}_{\vec k}(t_f,0)]_{11} &=& \eta_k(t_f) e^{-i \xi_{1\vec k}(t_f)},
\non \\
\cos \phi^{(0)}_{\vec k}(t_f) &=&\eta_k(t_f) \cos[\xi_{1 \vec
k}(t_f)]. \label{pbreg1} \eea Note that for the phase bands to
cross, {\it i.e.}, for $\phi^{(0)}_{\vec k}(t_f) = p \pi$, we
necessarily require a perfect overlap between the instantaneous and
initial ground state wave functions since we need $\eta_{\vec
k}(t_f)=1$ at the crossing time.

Next, in region $2$, between the first and the second avoided crossings, the
evolution operator in the adiabatic basis can be read off from
Eq.\ \eqref{uadrel1} as
\bea \vec c^{(2)}_{\vec k} (t) &=& U^{\rm ad}_{\vec k}(t, t_{1\vec k})
S_{1\vec k} U^{\rm ad}_{\vec k}(t_{1\vec k},0) \left(
\begin{array}{c} 1 \\ 0
\end{array} \right) \non \\
&=& \left(\begin{array}{c} e^{-i (\xi_{1\vec k}+
\xi_{2\vec k}(t_f) + \phi_{1 \vec k})}\sqrt{1-p_{1\vec k}} \\
-\sqrt{p_{1\vec k}} e^{-i (\xi_{1 \vec k} - \xi_{2 \vec k}(t_f)) }
\end{array} \right). \label{adregion2} \eea
Using Eqs.\ \eqref{adregion2}, \eqref{etadef} and \eqref{phasebands1}, some
straightforward algebra yields, for $t_{1 \vec k} \le t \le t_{2 \vec k}$,
\bea && [U^{(2)}_{\vec k}(t,0)]_{11} = \eta_k (t) \sqrt{1-p_{1\vec k}}
e^{-i (\xi_{1\vec k} + \xi_{2 \vec k}(t_f) + \phi_{1\vec k})} \non \\
&& - \sqrt{p_{1\vec k}(1-\eta_{\vec k} (t)^2)} e^{-i (\xi_{1 \vec
k}- \xi_{\vec k}(t_f))}, \label{pbreg2}\\
&&\cos \phi^{(1)}_{\vec k} (t) = \eta_k (t) \sqrt{1-p_{1\vec k}}
~\cos[
\xi_{1 \vec k} + \xi_{2\vec k}(t_f) + \phi_{1\vec k}] \non \\
&& ~~~~~~~~~~~~~~~~~~~~~~- \sqrt{p_{1\vec k}(1-\eta^2_{\vec k} (t))}~
\cos[\xi_{1\vec k} - \xi_{2\vec k}(t_f)]. \non \eea

Next, we discuss the situation in region $3$ for $t_{3 \vec k} \le t
\le t_{2 \vec k}$ for which there are two prior avoided level
crossings. To obtain the expressions for the phase bands, we note
that the wave function after two such level
crossings must come back to itself if $p_{\vec k}=1$ for each
crossing. Also we use a simplified notation where we denote
$\xi_{\vec k}(t_{1(2) \vec k},0) \equiv \xi_{1(2) \vec k}$, $
\xi_{\vec k}(t_f, t_{2\vec k}) \equiv \xi_{3 \vec k}(t_f)$,
$\xi^s_{\vec k}(t_f) = \sum_{i=1,2} \xi_{i \vec k} + \xi_{3 \vec
k}(t_f)$, and $\phi^s_{\vec k} = \sum_{i=1,2} \phi_{i, \vec k}$.
This yields, after some algebra,
\begin{widetext}
\bea \vec c^{(3)}_{\vec k}(t_f) &=& U^{\rm ad}_{\vec k}(t_f, t_{2\vec k})
S_2^T U^{\rm ad}_{\vec k}(t_{2 \vec k},t_{1\vec k})
S_{1\vec k} U^{\rm ad}_{\vec k}(t_{1\vec k},0) \left(
\begin{array}{c} 1 \\ 0
\end{array} \right) \non \\
&=& \left(\begin{array}{c} e^{-i (\xi^s_{\vec k}(t_f) + \phi_{\vec
k}^s)} \sqrt{(1-p_{1\vec k})(1-p_{2 \vec k})} + \sqrt{p_{1\vec k}
p_{2 \vec k}} e^{-i ( \xi_{\vec k}^s(t_f) - 2 \xi_{2 \vec k})} \\
\sqrt{p_{2\vec k}(1-p_{1 \vec k})} e^{-i (\phi_{1\vec k}+ \xi_{1
\vec k} + \xi_{2 \vec k} - \xi_{3 \vec k})} - \sqrt{p_{1\vec k}(1-p_{2
\vec k})} e^{i (\phi_{2 \vec k} + \xi_{2 \vec k} - \xi_{1 \vec k} + \xi_{3
\vec k}(t_f))} \end{array} \right). \label{adregion3} \eea
\end{widetext}
Using Eqs.\ \eqref{adregion3}, \eqref{etadef}, and \eqref{phasebands1},
we can obtain the expressions for the phase bands. The final result reads
\begin{widetext}
\bea &&\cos \phi^{(2)}_{\vec k}(t_f) = \eta_k(t_f)
\left[\sqrt{(1-p_{1\vec k}) (1-p_{2 \vec k})} ~\cos [\xi^s_{\vec
k}(t_f) + \phi_{\vec k}^s] + \sqrt{p_{1 \vec k} p_{2 \vec k}}
~\cos[\phi_{1\vec k}+ \xi_{1 \vec k} + \xi_{2 \vec k}
- \xi_{3 \vec k}(t_f)] \right] \non \\
&& + \sqrt{1-\eta_k^2 (t_f)} \left[\sqrt{p_{2\vec k}(1-p_{1 \vec k})}~
\cos[\phi_{1\vec k}+ \xi_{1 \vec k} + \xi_{2 \vec k} - \xi_{3 \vec k}(t_f)]-
\sqrt{p_{1\vec k}(1-p_{2 \vec k})} ~\cos[\phi_{2\vec k} + \xi_{2 \vec k} -
\xi_{1 \vec k} + \xi_{3 \vec k}(t_f)] \right]. \non \\
&& \label{pbreg3} \eea
\end{widetext}
Next, we consider the case where $p_{1 \vec k} = p_{2 \vec k} =
p_{\vec k} $ and $\phi_{1 \vec k} = \phi_{2 \vec k} = \phi_{\vec
k}$. Such equalities can be justified for high-symmetry points in
the graphene Brillouin zone as noted in Secs.\ \ref{secsymm} and \ref{secphd}.
In this case we obtain simpler expressions for the phase band given by
\begin{widetext}
\bea \cos(\phi^{(2)}(\vec k,t)) &=& \eta_{\vec k} (t) \left(
(1-p_{\vk})\cos(
\mu_{\vk}^s (t))+ p_{\vk} \cos(\mu_{\vk}^d (t))\right) \non \\
&& + \sqrt{p_{\vk}(1-p_{\vk})(1-\eta_{\vk}^2 (t))} ~\left( \cos(\mu^s_{\vk}
(t) - \phi_{s\vk})- \cos(\mu^d_{\vk} (t) - \phi_{s\vk}) \right), \non \\
\mu_{\vk}^s (t) &=& 2 \phi^s_{\vk} + \xi^s_{\vk}(t_f), \quad
\mu_{\vk}^d (t) = \xi_{1 \vk}-\xi_{2 \vk}+ \xi_{3 \vk}(t_f).
\label{pbreg3sim} \eea
\end{widetext}

For larger $n$ we can, in principle, write down the expressions for
$\cos(\phi^{(n)}(\vec k, t)$ following the same procedure. However,
these expressions get more complicated with increasing $n$. In what
follows, we concentrate on the expressions for the phase bands at
$t=t_f$ for which $\eta_{\vec k}(t_f)=1$ since, as argued in the
main text, the generic phase band crossings require this condition.
In this case, we first note that for $\eta_{\vec k}(t_f)=1$, after
$n$ avoided crossings, the matrix element of the evolution operator
$[U_{\vec k}(t_f,0)]_{11}$ operators is related to the wave function
$c^{(n+1)}_{1 \vec k}(t_f)$ in the adiabatic basis: $[U_{\vec
k}(t_f,0)]_{11}= c^{(n+1)}_{1 \vec k}(t_f)$. Next, we note that
after $n$ crossings, $\vec c^{\,(n+1)}_{\vec k}(t_f)$ is given by
\bea \vec c^{(n+1)}_{\vec k}(t_f) &=& U^{\rm ad}_{\vec k}(t_f,
t_{n-1\vec
k}) S_2^T \cdots \non \\
&& \times U^{\rm ad}_{\vec k}(t_{2 \vec k},t_{1\vec k})
S_{1\vec k} U^{\rm ad}_{\vec k}(t_{1\vec k},0) \left(
\begin{array}{c} 1 \\ 0
\end{array} \right). \eea
A few lines of algebra then lead to
\begin{widetext}
\bea c^{(2n+1)}_{1 \vec k}(t_f)&=& \sum_{jmax=0, 2, \cdots, 2n}
\sum_{\al} \prod_{j_{\al} = 1}^{jmax} (1-p_{j_{\al} \vec k})^{1/2}
\prod_{j'_{\al} \ne j_{\al}=1}^{2n-jmax} p_{j'_{\al} \vec k}^{1/2}
(-1)^{n_1} e^{-i\left[\phi_{\vec k}^s + \xi_{\vec k}^s(t_f) -\sum_a
(\ga_a \phi_{a \vec k} + \de_a \xi_{a \vec k}) \right]} ,
\label{coeffneq} \\
c^{(2n)}_{1 \vec k}(t_f) &=& \sum_{jmax=0}^{2n} \sum_{\al}
\prod_{j_{\al} = 1}^{jmax} (1-p_{j_{\al} \vec k})^{1/2}
\prod_{j'_{\al} \ne j_{\al}=1}^{2n-1-jmax} p_{j'_{\al} \vec k}^{1/2}
(-1)^{n_2} (1-\de_{n_{j'_{\al}},1}) e^{-i \left[\phi_{\vec k}^s +
\xi_{\vec k}^s(t_f) -\sum_a (\ga_a \phi_{a \vec k} + \de_a \xi_{a
\vec k}) \right]} , \non \eea
\end{widetext}
where the sum over the index $\al$ represents a sum over all
possible permutations of $j_{\al}$ and $j'_{\al}$ for a fixed
$jmax$, $n_1 = {\rm Max}[j'_{\al}]-{\rm Min}[j'_{\al}]+1$
provided ${\rm Min}[j'_{\al}] \ne 0$ and is $0$ otherwise,
$n_{j'_{\al}}$ denotes the number of occurrence of $j'$ ({\it
i.e.}, the number of $\sqrt{p_{j_{\al}\vec k}}$ factors) in a
permutation $\al$, and $n_2= n_{j_{\al}}$ with $ {\rm
Max}[j'_{\al}]<j_{\al}<{\rm Min}[j'_{\al}]$. The coefficients
$\ga_a$ and $\de_a$ for any given permutation $\al$ are given by
\bea \ga_a &=& 1 \, \, {\rm for}\, a \in j'_{\al}, \non \\
&=& 2 \, \, {\rm for}\, a \in j_{\al} \,{\rm with}\, j^{o
\prime}_{\al} < j_{\al} < j^{e \prime}_{\al}, \label{coeffnexp} \\
&=& 0 \, \, {\rm otherwise}, \non \\
\de_a &=& 2 \, \, {\rm for}\, a \in j_{\al},j'_{\al}\,\,
{\rm with} \, {\rm
Min}[j'_{\al}] < j_{\al},j'_{\al} < {\rm Max}[j'_{\al}] \non \\
&=& 0 \,\, {\rm if}\, a \in j_{\al},j'_{\al}\, {\rm with}\,
a-1 \in j'_{\al}\, {\rm and}\, \de_{a-1}^e =2, \non
\eea
where $j^{o \prime}_{\al}$ denotes any odd occurrence of $j'$
during a permutation, and $j^{e \prime}_{\al}$ denotes its next
occurrence in that permutation.

The expressions for the phase bands can then be obtained from Eqs.\
\eqref{coeffneq} by using $\cos[\phi^{(n)}(\vec k,t)]= {\rm
Re}[[U_{\vec k}(t_f,0)]_{11}]= {\rm Re}[c^{(n+1)}_{1 \vec k}(t_f)]$,
and leads to Eqs.\ \eqref{evencos1} and \eqref{oddcos1} in the main
text.

\vspace{.5cm} \centerline{\bf Acknowledgments} \vspace{.5cm}

D.S. thanks DST, India for Project No. SR/S2/JCB-44/2010 for
financial support. PM would like to thank Theoretical Physics
Department of IACS for hospitality and support. 


\vspace{.5cm}

\begin{thebibliography}{99}

\bib{rev1} J. Dziarmaga, Adv. Phys. {\bf 59}, 1063 (2010).

\bib{rev2} A. Polkovnikov, K. Sengupta, A. Silva, and M. Vengalattore,
Rev. Mod. Phys. {\bf 83}, 863 (2011).

\bib{rev3} A. Dutta, G. Aeppli, B. K. Chakrabarti, U. Divakaran, T. F.
Rosenbaum, and D. Sen, {\it Quantum phase transitions in transverse field spin
models: from statistical physics to quantum information} (Cambridge University
Press, Cambridge, 2015).

\bib{rev4} L. D'Alessio, Y. Kafri, A. Polkovnikov, and M. Rigol, Adv. Phys.
{\bf 65}, 239 (2016).

\bib{kz1} T. W. B. Kibble, J. Phys. A {\bf 9}, 1387 (1976); W. H. Zurek,
Nature (London) {\bf 317}, 505 (1985).

\bib{pol2} A. Polkovnikov, Phys. Rev. B {\bf 72}, 161201(R) (2005).

\bib{diva1} V. Mukherjee, U. Divakaran, A. Dutta, and D. Sen, Phys.
Rev. B {\bf 76}, 174303 (2007); C. De Grandi and A. Polkovnikov, in
{\it Quantum Quenching, Annealing, and Computation}, edited by A. K.
Chandra, A. Das, and B. K. Chakrabarti, Lecture Notes in Physics,
Vol. 802 (Springer, Heidelberg, 2010), p. 75.

\bib{ks1} K. Sengupta, D. Sen, and S. Mondal, Phys. Rev. Lett. {\bf 100},
077204 (2008).

\bib{ks2} D. Sen, K. Sengupta, and S. Mondal, Phys. Rev. Lett. {\bf 101},
016806 (2008).

\bib{pol3} A. Polkovnikov, Phys. Rev. Lett. {\bf 101}, 220402 (2008).

\bib{sondhi1} A. Chandran, A. Erez, S. S. Gubser, and S. L. Sondhi,
Phys. Rev. B {\bf 86}, 064304 (2012).

\bib{sdas1} S. R. Das, D. A. Galante, and R. C. Myers, \prl {\bf 112}, 171601
(2014); {\it ibid}, JHEP {\bf 1502}, 167 (2015).

\bib{sdas2} D. Das, S. R. Das, D. A. Galante, R. C. Myers, and K. Sengupta,
arXiv:1706.02322 (unpublished).

\bib{subir1} K. Sengupta, S. Powell, and S. Sachdev, Phys. Rev. A {\bf 69},
053616 (2004).

\bibitem{das1} A. Das, K. Sengupta, D. Sen, and B. K. Chakrabarti,
Phys. Rev. B {\bf 74}, 144423 (2006).

\bib{cala1} P. Calabrese and J. Cardy, J. of Stat. Mech. (2005) P04010;
{\it ibid.}, Phys. Rev. Lett. {\bf 96}, 136801 (2006).

\bibitem{silva1} A. Silva, Phys. Rev. Lett. {\bf 101}, 120603 (2008);
A. Gambassi and A. Silva, Phys. Rev. Lett. {\bf 109}, 250602 (2012);
S. Sotiriadis, A. Gambassi, and A. Silva, Phys. Rev. E {\bf 87}, 052129
(2013); P. Smacchia and A. Silva, Phys. Rev. E {\bf 88}, 042109 (2013).

\bibitem{dytr0} M. Heyl, A. Polkovnikov, and S. Kehrein, Phys. Rev. Lett.
{\bf 110}, 135704 (2013).

\bibitem{dytr1} C. Karrasch and D. Schuricht, Phys. Rev. B, {\bf 87}, 195104
(2013); N. Kriel, C. Karrasch, and S. Kehrein Phys. Rev. B {\bf 90}, 125106
(2014).

\bibitem{dytr2} F. Andraschko and J. Sirker, Phys. Rev. B {\bf 89}, 125120
(2014); E. Canovi, P. Werner, and M. Eckstein, Phys. Rev. Lett. {\bf 113},
265702 (2014); M. Heyl, Phys. Rev. Lett. {\bf 115}, 140602 (2015).

\bibitem{amit1} S. Sharma, U. Divakaran, A. Polkovnikov, and A. Dutta,
Phys. Rev. B {\bf 93}, 144306 (2016).

\bibitem{sau1} J. D. Sau and K. Sengupta, Phys. Rev. B {\bf 90}, 104306 (2014).

\bibitem{del1} A. del Campo and K. Sengupta, Eur. Phys. J. Special Topics
{\bf 224}, 189 (2015).

\bib{rg1} L. M. Sieberer, S. D. Huber, E. Altman, and S. Diehl, Phys. Rev.
Lett. {\bf 110}, 195301 (2013).

\bib {rg2} S. de Sarkar, R. Sensarma, and K. Sengupta, J. Phys. Condens.
Matter {\bf 23}, 305602 (2014).

\bib{rg3} A. Mitra and T. Giamarchi, Phys. Rev. Lett. {\bf 107}, 150602 (2011);
A. Mitra and T. Giamarchi, Phys. Rev. B {\bf 85}, 075117 (2012).

\bibitem{exp1a} For a review, see I. Bloch, J. Dalibard, and W. Zwerger,
Rev. Mod. Phys. {\bf 80}, 885 (2008).

\bibitem{exp1b} M. Greiner, O. Mandel, T. Esslinger, T.W. Hansch, and I. Bloch,
Nature (London) {\bf 415}, 39 (2002); W. S. Bakr, A. Peng, M. E. Tai, R. Ma,
J. Simon, J. I. Gillen, S. Folling, L. Pollet, and M. Greiner, Science
{\bf 329}, 547 (2010); J. Simon, W. Bakr, R. Ma, M. E. Tai, P. Preiss, and
M. Greiner, Nature (London) {\bf 472}, 307 (2011).

\bibitem{stu1} E. C. G. Stuckelberg, Helv. Phys. Acta {\bf 5}, 369 (1932).

\bibitem{stu2} W. D. Oliver, Y. Yu, J. C. Lee, K. K. Berggren, L. S. Levitov,
and T. P. Orlando, Science {\bf 310}, 1653 (2005); M. S. Rudner, A. V. Shytov,
L. S. Levitov, D. M. Berns, W. D. Oliver, S. O. Valenzuela, and T. P. Orlando,
Phys. Rev. Lett. {\bf 101}, 190502 (2008)

\bibitem{stu3} A. Dutta, A. Das, and K. Sengupta, Phys. Rev. E {\bf 92}, 012104
(2015).

\bibitem{asen1} A. Sen, S. Nandy, and K. Sengupta, Phys. Rev. B {\bf 94},
214301 (2016) .

\bibitem{adas1} A. Das, Phys. Rev. B {\bf 82}, 172402 (2010); S. S. Hegde,
H. Katiyar, T. S. Mahesh, and A. Das, Phys. Rev. B {\bf 90}, 174407 (2014).

\bibitem{sm1} S. Mondal, D. Pekker, and K. Sengupta, Europhys. Lett. {\bf 100},
60007 (2012); U. Divakaran and K. Sengupta, Phys. Rev. B {\bf 90}, 184303
(2014).

\bibitem{sk1} S. Kar, B. Mukherjee, and K. Sengupta, \prb {\bf 94}, 075130
(2016).

\bibitem{adas2} A. Lazarides, A. Das, and R. Moessner, \prl {\bf 112}, 150401
(2014); {\it ibid}, \pre {\bf 90}, 012110 (2014).

\bib{top1}M. S. Rudner, N. H. Lindner, E. Berg, and M. Levin, Phys. Rev.
X {\bf 3}, 031005 (2013); F. Nathan and M. S. Rudner, New J. Phys.
{\bf 17}, 125014 (2015); D. Carpentier, P. Delplace, M. Fruchart,
and K. Gawedzki, Phys. Rev. Lett. {\bf 114}, 106806 (2015).

\bib{top2} T. Kitagawa, E. Berg, M. Rudner, and E. Demler, Phys. Rev. B
{\bf 82}, 235114 (2010); Z. Gu, H. A. Fertig, D. P. Arovas, and A. Auerbach,
Phys. Rev. Lett. {\bf 107}, 216601 (2011); E. S. Morell and L.
E. F. Foa Torres, Phys. Rev. B {\bf 86}, 125449 (2012); M. Trif and
Y. Tserkovnyak, Phys. Rev. Lett. {\bf 109}, 257002 (2012); A.
Russomanno, A. Silva, and G. E. Santoro, Phys. Rev. Lett. {\bf 109},
257201 (2012); V. M. Bastidas, C. Emary, G. Schaller, and T.
Brandes, Phys. Rev. A {\bf 86}, 063627 (2012); V. M. Bastidas, C.
Emary, B. Regler, and T. Brandes, Phys. Rev. Lett. {\bf 108}, 043003
(2012); M. Tomka, A. Polkovnikov, and V. Gritsev, Phys. Rev. Lett.
{\bf 108}, 080404 (2012); A. Gom\'ez-Le\'on and G. Platero, Phys. Rev. B
{\bf 86}, 115318 (2012); A. Gom\'ez-Le\'on and G. Platero, Phys. Rev. Lett.
{\bf 110}, 200403 (2013); B. D\'ora, J. Cayssol, F. Simon, and R. Moessner,
Phys. Rev. Lett. {\bf 108}, 056602 (2012); D. E. Liu, A. Levchenko, and H. U.
Baranger, Phys. Rev. Lett. {\bf 111}, 047002 (2013); Q.-J. Tong,
J.-H. An, J. Gong, H.-G. Luo, and C. H. Oh, Phys. Rev. B {\bf 87},
201109(R) (2013); Y. T. Katan and D. Podolsky, Phys. Rev. Lett. {\bf
110}, 016802 (2013).

\bib{oka1} T. Kitagawa, T. Oka, A. Brataas, L. Fu, and E. Demler, Phys. Rev. B
{\bf 84}, 235108 (2011); T. Oka and H. Aoki, \prb {\bf 79}, 081406 (2009).

\bib{gil1} N. H. Lindner, G. Refael, and V. Galitski, Nature Phys. {\bf 7},
490 (2011); L. Jiang, T. Kitagawa, J. Alicea, A. R. Akhmerov, D. Pekker,
G. Refael, J. I. Cirac, E. Demler, M. D. Lukin, and P. Zoller, Phys. Rev.
Lett. {\bf 106}, 220402 (2011); N. H. Lindner, D. L. Bergman, G. Refael, and
V. Galitski, Phys. Rev. B {\bf 87}, 235131 (2013).

\bib{abhiskar1} M. Thakurathi, A. A. Patel, D. Sen, and A. Dutta, Phys. Rev. B
{\bf 88}, 155133 (2013); M. Thakurathi, K. Sengupta, and D. Sen, Phys. Rev. B
{\bf 89}, 235434 (2014).

\bib{bm1} B. Mukherjee, A. Sen, D. Sen, and K. Sengupta, \prb {\bf 94}, 155122
(2016).

\bib{kobo1} P. Weinberg, M. Bukov, L. D'Alessio, A. Polkovnikov, S. Vajna, and
M. Kolodrubetz, Phys. Rep. {\bf 688}, 1 (2017).

\bib{kundu1} A. Kundu and B. Seradjeh, Phys. Rev. Lett. {\bf 111}, 136402
(2013); A. Kundu, H. Fertig, and B. Seradjeh, \prl {\bf 113}, 236803 (2014).

\bib{nori1} S. N. Shevchenko, S. Ashhab, and F. Nori, Physics Reports
{\bf 492}, 1 (2010).

\bib{suzuki1} T. Fukui, Y. Hatsugai, and H. Suzuki, J. Phys. Soc. Jpn.
{\bf 74}, 1674 (2005).

\bib{degottardi} W. DeGottardi, M. Thakurathi, S. Vishveshwara and D. Sen,
Phys. Rev. B {\bf 88}, 165111 (2013).

\bib{saha} S. Saha, S. N. Sivarajan and D. Sen, Phys. Rev. B {\bf 95}, 174306
(2017).

\bibitem{exp2} L. P. Gavensky, G. Usaj, and C. A. Balseiro, Scientific
Reports {\bf 6}, 36577 (2016).

\bibitem{exp3} Y. H. Wang, H. Steinberg, P. Jarillo-Herrero P, and N. Gedik,
Science {\bf 342}, 453 (2013).

\bibitem{exp4} S. Mukherjee, A. Spracklen, M. Valiente, E. Andersson,
P. Ohberg, N. Goldman and R. R. Thomson, Nature Communications {\bf 8}, 13918
(2017).

\bibitem{exp5} J. W. McIver, D. Hsieh, H. Steinberg, P. Jarillo-Herrero, and
N. Gedik, Nature Nanotechnology {\bf 7}, 96 (2012).

\end{thebibliography}
\end{document}